*Article*

# A Multi-Agent, Laxity-Based Aggregation Strategy for Cost-Effective Electric Vehicle Charging and Local Transformer Overload Prevention

Kristoffer Christensen, Bo Nørregaard Jørgensen * and Zheng Grace Ma

SDU Center for Energy Informatics, Maersk Mc-Kinney Moeller Institute, The Faculty of Engineering, University of Southern Denmark, 5230 Odense, Denmark; kric@mmmi.sdu.dk (K.C.); zma@mmmi.sdu.dk (Z.G.M.)
* Correspondence: bnj@mmmi.sdu.dk

**Abstract:** The rapid electrification of transportation, driven by stringent decarbonization targets and supportive policies, poses significant challenges for distribution system operators (DSOs). When numerous electric vehicles (EVs) charge concurrently, local transformers risk overloading—a problem that current tariff-based strategies do not adequately address. This paper introduces an aggregator-based coordination mechanism that shifts EV charging from congested to underutilized periods using a rule-based scheduling algorithm. Unlike conventional methods that depend on complex real-time pricing signals or optimization-heavy solutions, the aggregator approach uses a simple yet effective "laxity" measure to prioritize charging flexibility. To assess technical and economic viability, a multi-agent simulation was developed to replicate residential user behavior and DSO constraints under the use of a 400 kVA low-voltage transformer. The results indicate that overloads are completely eliminated with minimal inconvenience to users, whose increased charging costs are offset by the aggregator at an annual total of under DKK 6000—significantly lower than the cost of infrastructure reinforcement. This study contributes by (i) quantifying the compensation needed to prevent large-scale overloads, (ii) presenting a replicable, computationally feasible, rule-based aggregator model for DSOs, and (iii) comparing aggregator solutions to costly transformer upgrades, underscoring the aggregator's role as a viable tool for future distribution systems.

## 1. Introduction

Efforts to reduce carbon emissions worldwide have accelerated the transition to electric vehicles (EVs) in the transportation sector. Policy targets, such as Denmark's aim for one million EVs by 2030, reflect this growing momentum [1,2]. While these initiatives promise environmental benefits, they also strain low-voltage distribution grids. Residential transformers, originally sized for household loads, increasingly risk overload when multiple EVs charge in overlapping time windows [3]. Even sophisticated charging algorithms that respond to aggregated electricity prices may inadvertently create synchronized demand peaks, ultimately failing to relieve strain on distribution assets [3].

Existing tariff designs exacerbate this challenge. Danish DSOs, for instance, must rely on transparent, nondiscriminatory, and market-based approaches [4] and cannot easily deploy highly dynamic price signals [5]. Thus, Danish distribution tariffs are currently restricted to time-of-use schemes that often remain static or insufficiently granular, leading to persistent local congestion when consumer responses cluster [3,6]. Upgrading transformers is technically straightforward but expensive and may be underutilized if load growth projections do not fully materialize. Recognizing these limitations, newer regulatory frameworks highlight local flexibility markets and aggregator-based load coordination as alternatives to costly infrastructure expansion [7,8].

Despite considerable advancements in EV demand response, three primary gaps hinder practical adoption. First, most aggregator studies focus on large-scale or regional grid services, with less attention paid to localized overloads at individual transformers [9–11]. Second, advanced optimization- or machine-learning-based schedulers (for example, those discussed in [12,13]) require extensive computational resources and forecasting precision, making them less accessible to DSOs seeking near-term deployment. Third, the overall cost-effectiveness of aggregator solutions compared to capital-intensive upgrades remains insufficiently quantified. Many approaches concentrate on minimizing EV users' electricity bills or maximizing aggregator profits [14,15], rather than exploring how DSOs might leverage a relatively simple aggregator model to preempt or resolve local transformer overload.

In response to these gaps, this paper introduces a novel aggregator-based coordination mechanism designed to (i) mitigate local transformer overload, (ii) minimize user inconvenience by compensating for any additional energy costs caused by load shifting, and (iii) evaluate cost-effectiveness against the alternative of infrastructure reinforcement. The aggregator algorithm applies a rule-based strategy that orders vehicles according to their "laxity", a measure of their scheduling flexibility [16]. Compared with advanced optimization or machine learning approaches [17–19], the proposed rule-based method aims for a balance of simplicity, transparency, and operational speed—factors critical for DSOs.

Two major contributions define the novelty of this research. First, the paper quantitatively establishes the minimum compensation cost required to achieve zero overloads for a 400 kVA transformer with full EV adoption, demonstrating its minimal value relative to the expenditure of upgrading or replacing a transformer station. Second, it integrates a multi-agent simulation model with a detailed business ecosystem perspective [20,21], thereby capturing the micro-level interactions among EV owners, the aggregator, and DSOs. By simulating each agent's behavior over an entire year, the study provides robust insights into emergent charging patterns and overload episodes in real-world-like conditions.

To investigate the aggregator's performance, this study adopts a multi-agent-based simulation, wherein each household, EV, and system operator acts autonomously, yet interacts according to market and regulatory rules. The aggregator modifies individual EV schedules only when it detects a risk of overload, compensating users for higher electricity prices in shifted hours. This design treats the aggregator as a practical, cost-focused mechanism rather than a profit-maximizing entity. The rule-based model avoids heavy computational overhead and extends more naturally to real-time operations, making it promising for DSOs needing immediate solutions. The multi-agent platform is particularly well suited to evaluating emergent behavior, as local congestion events often arise unexpectedly when many agents follow similar price signals.

The remainder of this paper unfolds as follows. The Section 2 presents a detailed review of the relevant literature on EV home charging aggregation, highlighting state-of-the-art methods, local flexibility market concepts, and rule-based coordination. The subsequent Section 3 explains how the multi-agent simulation is constructed, including the aggregator's role and operational logic. This paper then introduces the case study—a Danish low-voltage grid—and outlines the baseline and aggregator scenarios for the simulation experiments. Following this, the results are presented, focusing on overload mitigation, cost impacts, and sensitivity to charging behaviors. The Section 6 interprets these findings in the context of the existing research, addressing both theoretical implications and practical considerations for DSOs. Finally, the conclusion summarizes key insights, clarifies the study's limitations, and suggests directions for future research, especially regarding partial aggregator participation and integration with additional distributed energy resources.

## 2. Literature Review

This section reviews the existing literature on EV aggregation models, including optimization-based, AI-driven, and rule-based strategies, and highlights their applicability to managing distribution grid congestion. Additionally, it explores local flexibility markets as a potential framework for integrating EV aggregation solutions. By identifying gaps in the existing research, this review sets the foundation for the proposed aggregator-based coordination mechanism.

### 2.1. EV Aggregation in Distribution Grids

The rapid proliferation of EVs is transforming distribution networks, creating both challenges and opportunities for enhancing grid stability. Government incentives worldwide have accelerated EV adoption, prompting a critical need for solutions that address localized stresses on distribution infrastructure. Recent work increasingly views aggregated EV fleets as essential resources in maintaining grid reliability, particularly when integrated into concepts such as Virtual Power Plants (VPPs) and Vehicle-to-Grid (V2G) systems. Studies have shown that distributed energy resources (DERs), including EVs, can provide valuable load management and ancillary services when aggregated [9–11]. However, a key challenge lies in the inherent volatility and intermittency of DERs, which demands sophisticated tools for ultra-short-term forecasting and real-time coordination [9].

In these aggregated configurations, the aggregator is a pivotal actor, pooling diverse DERs—ranging from EVs to household batteries—into a unified portfolio whose collective output can be optimized. By centralizing load control, aggregators facilitate reliable market participation and enhance grid support [22–24]. The specific contribution of EVs, when viewed as flexible controllable loads, has been underscored by research showing that well-managed charging can improve microgrid operations [25], reduce demand peaks via V2G interfaces [26], and prevent undesirable impacts on both transmission and distribution systems [27]. Additional investigations highlight the benefits of synchronizing EV charging with renewable energy availability [28,29]. Despite these advantages, infrastructure capacity constraints remain a concern, particularly in dense urban networks where large-scale EV adoption may exceed existing distribution limits [30,31]. Consequently, the literature increasingly underscores the need for specialized aggregation approaches tailored to residential distribution systems.

### 2.2. EV Aggregation Methodologies and Technologies

The methodology underpinning EV aggregation has evolved significantly, reflecting diverse grid requirements and technological capabilities. Approaches can generally be categorized into V2G- and DER-focused frameworks, optimization-based methods, AI-driven solutions, and rule-based strategies.

- V2G and DER Integration

V2G technology has captured considerable attention due to its potential for two-way energy flow between EVs and the grid. As shown by [32], EVs operating under a V2G model can bolster resilience by serving as DERs during periods of high demand. Another stream of research integrates V2G capability with reactive power management, ensuring both active and reactive power support to maintain voltage stability and power quality [33]. While these dual-sided interactions open new avenues for system flexibility, they also necessitate careful scheduling and robust infrastructure to handle bi-directional power flows.

- Optimization-Based Approaches

Optimization-based EV aggregation models typically use mathematical or heuristic algorithms to schedule charging under constrained objectives, such as minimizing costs or mitigating grid congestion. For example, Ref. [34] introduced a two-level hierarchical optimization model that balances user satisfaction with grid stability, which contrasts with more traditional, grid-centric strategies that disregard the consumer experience. A hybrid coordination scheme that combines centralized and decentralized logic is proposed by [35], offering improved operational efficiency by adapting to real-time grid and consumer conditions.

Further refining these concepts, Ref. [14] demonstrated that EV aggregators could successfully participate in ancillary service markets when supported by appropriate optimization frameworks, while Refs. [15,36] used bi-level optimization to manage charging schedules while providing backup energy during peak demand. Although such approaches can be highly effective, they may be constrained by the computational intensity of solving non-linear, large-scale problems, as well as by the need for the accurate, high-resolution forecasting of EV availability and demand.

- AI-Based Methods

Recent advances in artificial intelligence have also influenced EV aggregation strategies. Reinforcement learning has proven particularly promising: Ref. [12] described a framework that autonomously adjusts charging schedules based on real-time system conditions, while Ref. [13] introduced a model-free reinforcement learning method that adapts robustly to uncertainties in EV user behavior. Research on decentralized energy management for urban EV hubs further underscores the value of distributed control mechanisms, which can reduce the complexity of centralized coordination [37]. These AI-based methods, however, must often contend with large data requirements, ensuring robust communication links and significant training or run-time computational demands.

- Rule-Based Strategies

Despite the sophistication of optimization-based and AI-driven approaches, rule-based strategies retain their appeal for EV aggregation, particularly in real-time applications with limited computational resources. These strategies rely on predefined heuristic rules, which ensure a deterministic structure and faster decision-making. According to Ref. [38], rule-based algorithms are widely adopted in energy management systems due to their ease of implementation, robustness, and lower computational overhead. A rule-based energy management system, combined with proportional–integral control, can dynamically manage energy flows in complex systems [39]. Comparative studies have shown that while optimization-based methods may yield higher overall efficiency, rule-based approaches are preferred when real-time responsiveness and low computational requirements are paramount [40,41]. Yet, rule-based solutions can lack adaptability under highly variable conditions [42]. Notably, Refs. [14,43] illustrate that such methods can support ancillary services and reserve management when carefully tuned, offering a practical alternative that balances simplicity with operational effectiveness. Furthermore, when additional factors such as battery aging are incorporated into the system design, as demonstrated in [44], the complexity of rule-based approaches increases significantly. Unlike optimization-based methods, which can adjust to new constraints through iterative solutions, rule-based approaches require manual adjustments to accommodate evolving operational demands, making their scalability and adaptability more challenging.

*2.3. Local Flexibility Markets for EV Aggregation*

The emergence of local flexibility markets (LFMs) further expands the avenues for EV aggregation, enabling the trade of flexibility services among various stakeholders in a localized setting. Researchers have stressed the importance of regulatory frameworks that encourage fair compensation for aggregators, thereby incentivizing broader participation [45]. Transmission system operators and DSOs play a critical role in estimating flexibility demand and ensuring that localized congestion and voltage concerns are well managed [46].

Market-based instruments often require advanced metering, robust communication infrastructures, and clear definitions of how local and wholesale markets interact. The EcoGrid 2.0 project, analyzed by [47], provides insight into the potential for real-time market signals to reduce peak loads, while Ref. [48] underlines the technical challenges posed by metering and communication constraints. Crucially, accurate price signals and effective participation incentives have proven essential to achieving cost-effective flexibility procurement [49,50]. Nevertheless, fully integrating LFMs into residential EV charging ecosystems is complicated by coordination challenges, variable consumer behavior, and the practicality of real-time bidding. For these reasons, the present study employs a rule-based load-shifting approach—an arguably simpler mechanism—to investigate how DSOs can reliably mitigate transformer overload while compensating participants at minimal cost.

*2.4. Practical Implementation and Maturity of EV Aggregation*

EV aggregation is progressing through various stages of maturity, from pilot trials to commercial-scale deployments. Early initiatives often focused on validating the feasibility of V2G operations, as exemplified by the University of Delaware's collaboration with Delmarva Power, which used aggregated Ford Mach-E EVs for grid services under emerging regulatory conditions [51]. OVO Energy's pilot in the UK similarly integrated Nissan LEAFs into frequency regulation markets via bidirectional chargers [52].

More recent projects signal the commercial viability of large-scale EV aggregation. The Guadalupe Valley Electric Cooperative and Tesla's partnership in Texas showcased the integration of residential battery storage with EVs in the ERCOT Aggregated Distributed Energy Resource (ADER) program [53]. A related effort at Bandera Electric Cooperative employed VPP technology to coordinate residential solar and EV charging with the aim of peak reduction and enhanced grid resilience [54]. Octopus Energy's use of intelligent charging and discharging mechanisms further illustrates the commercial application of EVs as flexible loads in VPP environments [55]. Project Sciurus, the largest residential V2G trial worldwide, aggregates over 300 charging points and participates in the UK's Piclo demand–response market, offering compensation for each kilowatt-hour exported to the grid [56–58]. While these developments demonstrate the technical feasibility and economic promise of EV aggregation, widespread adoption remains limited by regulatory barriers, interoperability issues, and cost considerations in deploying new hardware.

*2.5. Identified Limitations and Gaps in the Literature*

Despite the diverse range of studies that investigate EV aggregation, several gaps persist. One notable shortcoming is the limited examination of local aggregation as a means of alleviating overload at residential transformers, even though high EV penetration can impose severe stresses at the distribution level [30,31]. Although a variety of aggregator roles and large-scale EV coordination models have been proposed, few studies have explicitly addressed the aggregator's function in residential charging ecosystems or investigated the optimal strategies for integrating additional resources, such as photovoltaic systems, home batteries, and heat pumps, within a unified energy management framework.

Another unresolved issue involves balancing economic and operational demands. Many studies optimize grid benefits while overlooking the additional costs borne by EV owners, raising questions about fairness and consumer acceptance. Cost trade-offs remain insufficiently quantified. Furthermore, although advanced optimization and AI-based approaches can be highly effective, their real-world deployment is often constrained by computational burden and forecasting requirements. Rule-based methods offer a potential alternative but need further validation, particularly in dynamic residential scenarios. Finally, international standard compliance—including ISO 15118 and IEC 61850—is rarely addressed in simulation-based studies, even though these standards are crucial for achieving large-scale, interoperable deployments.

Building on the existing body of work, the present research investigates the efficacy of a localized, rule-based aggregator approach in mitigating transformer overload within a residential distribution network. This focus directly addresses the limited exploration of EV home charging aggregation, provides data-driven insights into consumer cost implications, and paves the way for the future integration of multiple DERs in a unified energy management system. By drawing attention to the practicality of simple, rapidly deployable strategies, this study also serves as a bridge between theoretical optimization models and the complex realities DSOs face in managing distribution grid constraints.

The proposed aggregator-based coordination method introduces several key innovations that distinguish it from previous studies. First, it employs a novel rule-based strategy centered on a "laxity" measure, which quantifies each EV's scheduling flexibility. This approach contrasts sharply with conventional optimization-based or AI-driven models that rely on complex, computationally intensive algorithms and precise forecasting. By leveraging locally available information, our method achieves rapid, deterministic decision-making, making it highly suitable for real-time DSO operations. Second, the method explicitly focuses on mitigating transformer overload—a critical and often under-addressed issue in residential distribution systems—by quantifying the minimum compensation required to prevent costly grid reinforcements. Finally, the integration of a multi-agent simulation framework with a detailed business ecosystem model offers a dual-level analysis: it captures both micro-level interactions among individual EV users and macro-level grid dynamics. This holistic approach not only improves operational feasibility but also bridges the gap between theoretical control strategies and practical, cost-effective solutions for DSOs.

## 3. Methodology

This study focuses on a representative suburban low-voltage distribution network where clustered EV charging threatens to overload residential transformers. Such application scenarios are increasingly common in regions with high EV adoption and the possibility to charge at home, making localized coordination strategies both technically necessary and economically relevant. Specifically, this study investigates how an aggregator-based coordination mechanism can mitigate transformer overloads in residential EV charging. To analyze these effects, the research adopted a multi-agent simulation approach augmented by a business model development and modeling method proposed by [20]. The overarching aim is to capture the roles, interactions, and behaviors of diverse stakeholders—such as DSOs, EV users, charging equipment operators, and data providers—within a unified framework. This approach enables an in-depth exploration of emergent phenomena in EV home charging under both decentralized and centralized charging control strategies.

The prior literature has explored multiple methods for coordinating EV charging, including purely decentralized tariff-based strategies, centralized optimization algorithms, and sophisticated machine-learning models. Price-based decentralized strategies rely on price signals, such as time-of-use or real-time prices, but may lead to synchronized charging spikes and insufficient load diversification. Optimization approaches can be effective but often involve high computational overhead and rely heavily on accurate forecasting. Machine learning techniques, while powerful, may require extensive training data and may be less transparent or more difficult to implement for real-time applications (see discussion in [12,13]). By contrast, a rule-based agent framework offers lower complexity, clear decision paths, and rapid deployment—especially suited for DSOs that must react quickly to local overload conditions. Given these trade-offs, a multi-agent simulation with a rule-based aggregator was deemed most appropriate to test how localized load shifting can address transformer constraints in a cost-effective, computationally tractable manner.

The chosen methodology integrates two primary elements: a business ecosystem modeling method [20,21] and agent-based simulation [59–62]. This dual-layer approach was justified by the need to explore both the techno-economic interactions among grid participants and the emergent, decentralized behaviors of individual EV users. The ecosystem mapping technique allows each actor, object, and interaction to be clearly defined, thus aligning with the study's objectives of modeling real-world relationships and transactions within EV charging. The agent-based method operationalizes this ecosystem by simulating behaviors at the micro-level (e.g., individual EV charging sessions) and assessing the macro-level outcomes (e.g., transformer loading). Similar multi-agent frameworks have demonstrated strong applicability for evaluating DER and EV integration scenarios [3,63]. Hence, the synergy between business ecosystem modeling and agent-based simulation is uniquely suited to capture the interplay of technical, economic, and social factors in EV home charging aggregation.

While alternative approaches can deliver high precision under controlled conditions, they often overlook the dynamics and diverse perspectives of stakeholders within the broader business ecosystem. In contrast, a multi-agent-based simulation that models the entire business ecosystem provides a transparent, flexible, and scalable framework. This framework is particularly well suited to capture the dynamic, complex interactions that DSOs face in real-world environments. Although this approach requires more detailed calibration and validation, its ability to replicate the varied interactions among stakeholders makes it an ideal tool for assessing the practical implementation of EV aggregation strategies.

Below is a detailed description of the agent-based model and aggregator logic.

### 3.1. Ecosystem Mapping with Extended Aggregator Role

Figure 1 illustrates the extended EV home charging ecosystem, adapted from [21]. The new aggregator role, managed by the DSO, was added to the existing landscape of actors (domestic consumers, data hubs, meter operators) and their interactions. Table 1 shows the additional interactions involved in enabling the aggregator function, including the exchange of charging schedules, compensation costs, and baseload predictions.

**Table 1**. EV home charging ecosystem extension with aggregator role.

| Interaction Direction from: | | Interaction Properties | | Interaction Direction to | |
|---|---|---|---|---|---|
| Actor/Object | Role | Interaction Content | Interaction Type | Actor/Object | Role |
| DSO | Aggregator | Send new charging schedule | Information | Charging box | Charging control |
|  |  | Send compensation cost | Monetary | Domestic consumer | Electric vehicle user |
|  | Meter operator | Send baseload (prediction) |  |  |  |
| Charging box | Charging control | Send charging schedule | Information | DSO | Aggregator |
| Data hub | Data hub | Send day-ahead prices |  |  |  |

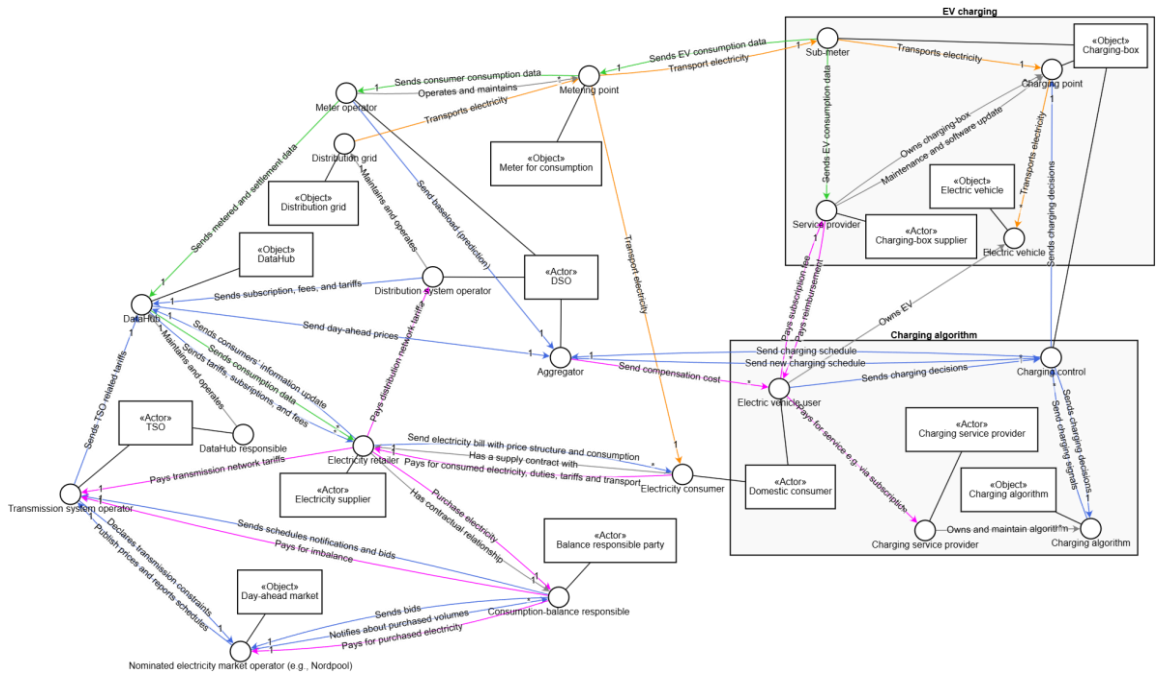

**Figure 1.** Ecosystem mapping of the EV home charging ecosystem with decentralized charging algorithms and an aggregator (adapted from [21]). 1 = interaction with one and * = interaction with many (>1).

Figure 1 depicts these newly introduced connections (in pink, blue, green, orange, and grey) within the broader EV charging ecosystem.

*3.2. Agent-Based Simulation Platform*

Agent-based modeling was employed to replicate the EV home charging ecosystem, incorporating both low-level individual behaviors (e.g., EV user charging decisions) and macro-level emergent outcomes (e.g., transformer load profiles). AnyLogic was chosen as the modeling tool because it supports multiple simulation paradigms—agent-based, discrete-event, and system dynamics—and has a proven track record in energy sector applications [59–62,64]. Drawing on [3,63], the following functions and interfaces are central to the aggregator's operation:

- addPredictedBaseLoad: This method is used by each domestic consumer agent once every 24 h to provide a baseload forecast. In practice, the baseload is implemented in the domestic consumer agent, and these updates enable the aggregator to incorporate the expected non-EV consumption into its overload calculations.
- addCustomer: This is used at the start of the simulation to register all EV users under the aggregator's purview.
- addFlexOffer: This is invoked by the charging control role whenever an EV plugs in. It transmits a proposed charging schedule (including time slots and power levels), along with critical information such as the planned departure time. This "FlexOffer" object represents the user's initial, possibly automated, charging strategy.
- unplugEV: This is triggered by the charging control role to remove the corresponding EV from the aggregator's scheduling portfolio upon unplugging.

Figure 2 presents the interfaces for the aggregator role and the necessary extension to the existing charging control interface. The aggregator must receive and process charging schedules, baseload forecasts, and user constraints. It can then recalculate and dispatch updated schedules to participating EV users.

The modularity in this design ensures that future expansions, such as integrating photovoltaic systems or home batteries, can follow the same interface pattern. Each new component simply adopts the agent role principle and connects with the existing methods accordingly.

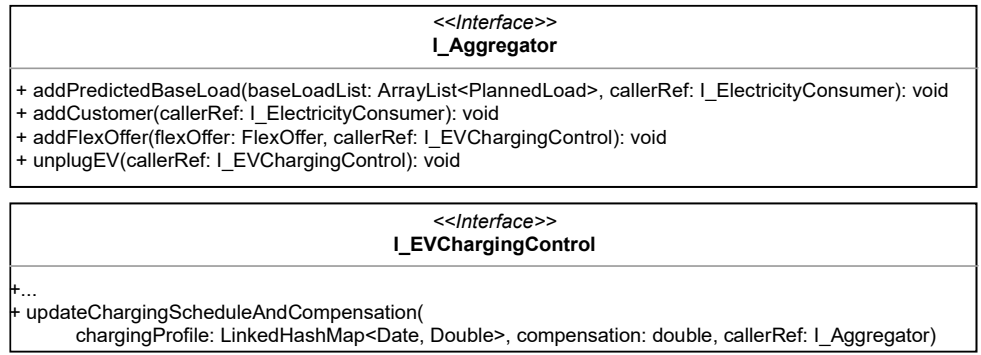

**Figure 2.** Interfaces of the aggregator role and the addition to the charging control interface.

*3.3. Rule-Based Decision Algorithm for EV Aggregation*

This study used a rule-based algorithm with a "laxity" measure to prioritize EVs with the most flexible schedules, enabling rescheduling with minimal user impact. Unlike optimization-based and AI-driven methods, which require high computational resources and precise forecasting. This section details the logic and implementation of the rule-based aggregator model, including overload detection, scheduling prioritization, and compensation mechanisms. By focusing on a simple yet effective scheduling approach, the algorithm ensures computational efficiency while delivering tangible grid benefits, making it a viable solution for DSOs that face increasing EV adoption.

3.3.1. Centralized Load-Shifting Concept

The primary aim of the aggregator is to prevent transformer overload by rescheduling EV charging sessions. The aggregator is assumed to be controlled by the DSO, which grants it the authority to modify individual charging plans when needed. This centralized strategy overrides the decentralized real-time pricing (RTP) behaviors of individual EV users if the projected load threatens to exceed the transformer capacity. The basic workflow is summarized in Figure 3.

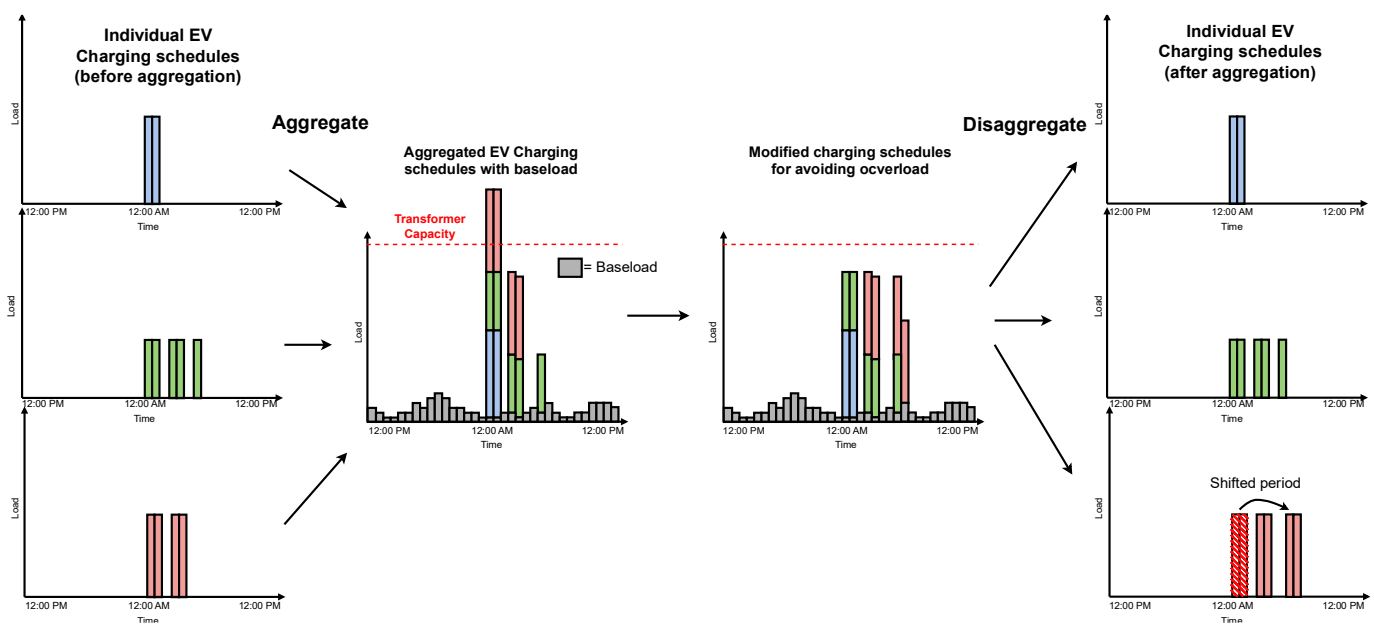

**Figure 3**. Logic of load shifting to avoid overload using aggregation.

3.3.2. Implementation Steps

When overload is detected, the aggregator employs a rule-based approach that measures each EV's flexibility via a concept termed "laxity". Higher laxity indicates that an EV has a larger buffer between charging completion time and the departure time, implying greater scheduling flexibility. The aggregator targets EVs with the highest laxity first, assuming these users can be shifted at minimal inconvenience. If no overload is present, the aggregator remains dormant to avoid unnecessary interventions.

Figure 4 details the step-by-step flow of the algorithm. The numbered sequence can be summarized as follows. First, the aggregator receives a baseload forecast from consumers, updated every 24 h. Once an EV is plugged in, the charging box (charging control role) sends a "FlexOffer" object to the aggregator, which contains the starting time, expected unplugging time, and a preliminary charging schedule (see Figure 5 for the class diagram). The aggregator aggregates all FlexOffers with the existing baseload to estimate transformer usage. If the combined forecasted load does not exceed capacity, no action is taken. Otherwise, the aggregator pinpoints the time intervals of expected overload and compiles a list of EVs charging in those intervals. The algorithm identifies the EV with the highest laxity, zeroes out the power in the overload period, and searches for alternative time slots within the same or subsequent cheapest hours (Figure 6 illustrates a typical load-shifting example within an hour). The aggregator updates the schedule in a backward-search manner, ensuring minimal cost impact if possible. Once rescheduling is complete, the aggregator sends updated schedules and compensates the EV users for any additional cost incurred, thereby preserving economic neutrality from the user's perspective.

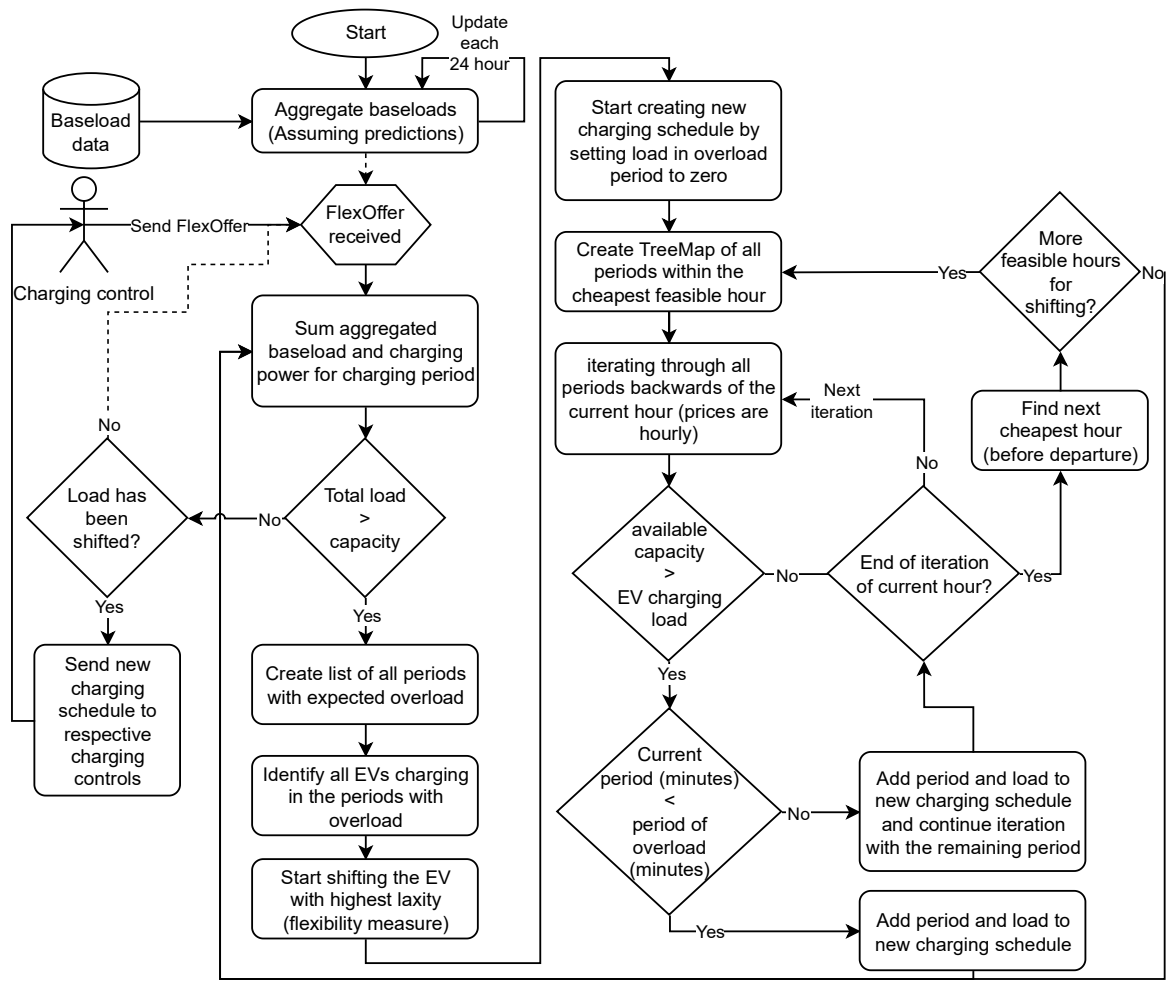

**Figure 4.** Flow chart of aggregator logic.

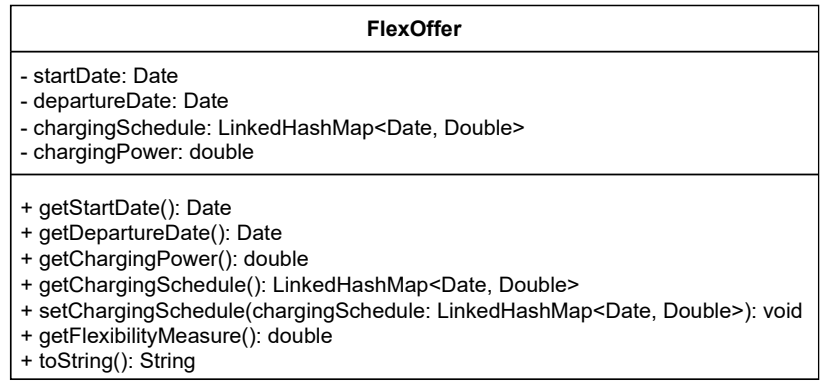

**Figure 5.** FlexOffer class diagram.

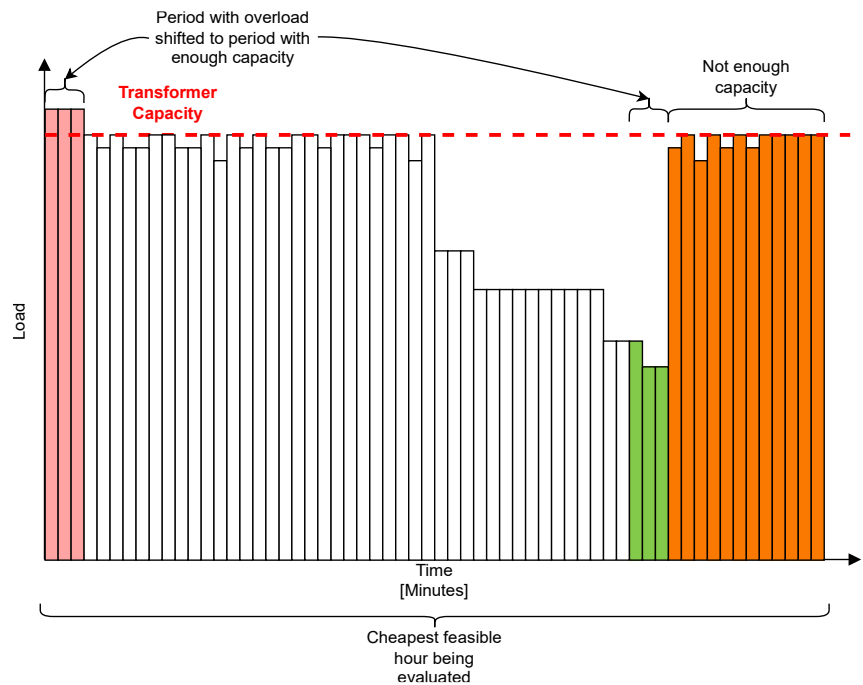

**Figure 6.** An example illustrating how a period with overload is shifted within the same hour.

A pseudo-code implementation of this procedure is shown in Algorithm 1. The code iterates through overload periods, sorts EVs by laxity, and systematically shifts charging load to available lower-load time slots, verifying capacity and departure constraints before finalizing adjustments.

---

**Algorithm 1. EV Aggregator Load Shifting Algorithm**

---

**Require:** Baseload forecast $B_t$, FlexOffers $F=\{F_1,F_2,\ldots,F_N\}$, maximum capacity $C_{max}$

**Ensure:** Adjusted EV charging schedules $S'$

1. Aggregate $B_t$ and $F$ to form expected load profile $L_t$
2. **if** no overload detected **then**
3. 　　Wait for new FlexOffers
4. **else**
5. 　　Identify periods $P_{overload}$ where $L_t > C_{max}$
6. 　　**for** each $P_{overload}$ **do**
7. 　　　　Create list of EVs charging in $P_{overload}$
8. 　　　　Sort EVs by highest laxity

| | |
|---|---|
| 9. | **for** each EV *e* in list **do** |
| 10. | 　Set charging power to zero in $P_{overload}$ |
| 11. | 　**for** each available period in the same hour (starting from end) **do** |
| 12. | 　　**if** sufficient capacity available and conditions satisfied **then** |
| 13. | 　　　Allocate charging to this period |
| 14. | 　　　**Break** |
| 15. | 　　**end if** |
| 16. | 　**end for** |
| 17. | 　**if** no valid period found **then** |
| 18. | 　　Search in the next cheapest hour before departure |
| 19. | 　　**if** no available periods remain **then** |
| 20. | 　　　Keep EV off during $P_{overload}$ |
| 21. | 　　**end if** |
| 22. | 　**end if** |
| 23. | **end for** |
| 24. | **end for** |
| 25. | Update load profile and check for remaining overloads |
| 26. | **if** overloads remain **then** |
| 27. | 　Repeat process until resolved or shifting is impossible |
| 28. | **end if** |
| 29. | Send updated schedules *S'* and compensation costs to charging controllers |
| 30. **end if** | |

The pseudo-code in Algorithm 1 illustrates how the aggregator functions:

1. The aggregator receives the baseload forecast.
2. An EV plugs in, and the aggregator receives a FlexOffer specifying start time, departure time, and an initial schedule.
3. All FlexOffers are combined to form a preliminary load forecast.
4. If the forecast exceeds capacity, the aggregator identifies the affected periods.
5. The aggregator generates a list of EVs scheduled during those overloads.
6. The EV with the highest laxity is chosen first.
7. The aggregator zeroes out power in the overload window, temporarily pausing charging.
8. The aggregator checks for free capacity within the same or other hours prior to departure, moving the required charging period to those free slots.
9. If no suitable slots can be found, the EV's charging remains paused until the next iteration.
10. Once all overloads are resolved, updated schedules and compensation costs are sent to affected EVs.

*3.4. Key Assumptions*

The proposed simulation framework is built upon several assumptions, which are necessary to model complex grid–user interactions in a tractable and representative manner. The most critical assumptions are listed below:

- User behavior is consistent and follows historical patterns. Charging schedules are based on empirical load profiles and assume users do not override aggregator control signals.

- EVs are driving each day. EVs will have a period each day of not being available for aggregation.
- All EV owners participate in the aggregator scheme. Partial or voluntary participation is not modeled, allowing a clearer assessment of theoretical upper-bound performance.
- Perfect hourly baseload predictions. Small load fluctuations within an hour are neglected and averaged for the hour due to available data. Furthermore, the load predictions are used to reschedule the charging schedules and are simulated as fully accurate.
- No physical grid constraints beyond transformer overloads are modeled. Voltage fluctuations, phase imbalances, and power quality issues are excluded from the current scope.
- Compensation payments are sufficient to incentivize user compliance. Users are assumed to accept schedule shifts as long as their costs are covered.

These assumptions are considered appropriate for an exploratory assessment of aggregator coordination potential. For example, the assumption that EVs are driven daily limits the aggregator's rescheduling flexibility, helping to identify user impacts under these more constrained conditions. However, the distance driven still reflects realistic behavior, as it is based on statistically derived probability distributions. Modeling full participation allows for the estimation of best-case outcomes. Likewise, excluding voltage issues is justified given the study's focus on transformer-level loading constraints, which are a primary bottleneck in many suburban networks. Nonetheless, these assumptions also introduce limitations. In practice, incomplete user participation or inaccurate load prediction could reduce aggregator effectiveness. Future works should investigate mixed participation models that are closer to real-life user behavior and broader grid constraints to validate the robustness of the proposed approach under more variable conditions.

## 4. Case Study and Scenario Design

To validate the proposed aggregator-based charging coordination strategy, a case study was conducted using real-world residential grid data from Strib, Denmark. The selected distribution network consists of 126 households connected to a 400 kVA transformer, reflecting a typical low-voltage grid facing potential overload due to high EV adoption. Two scenarios were analyzed: (i) a baseline scenario where EV users charge based on real-time electricity pricing, and (ii) an aggregation scenario where the aggregator actively shifts charging loads to avoid transformer stress. Key parameters such as household consumption profiles, EV charging behaviors, electricity tariffs, and market conditions were integrated into the simulation. This section outlines the experimental setup, data sources, and performance metrics used to compare the effectiveness of decentralized versus aggregator-based load management strategies.

### 4.1. Distribution Network and Household Characteristics

This study applied the outlined methodology to a radial distribution network in the city of Strib, Denmark, as shown in Figure 7. The network comprises 126 residential consumers connected below a 10/0.4 kV transformer rated at 400 kVA, with each household limited to three phases of 25 amps (17.3 kVA). Figure 7 illustrates the geographic layout of the distribution lines and household connections.

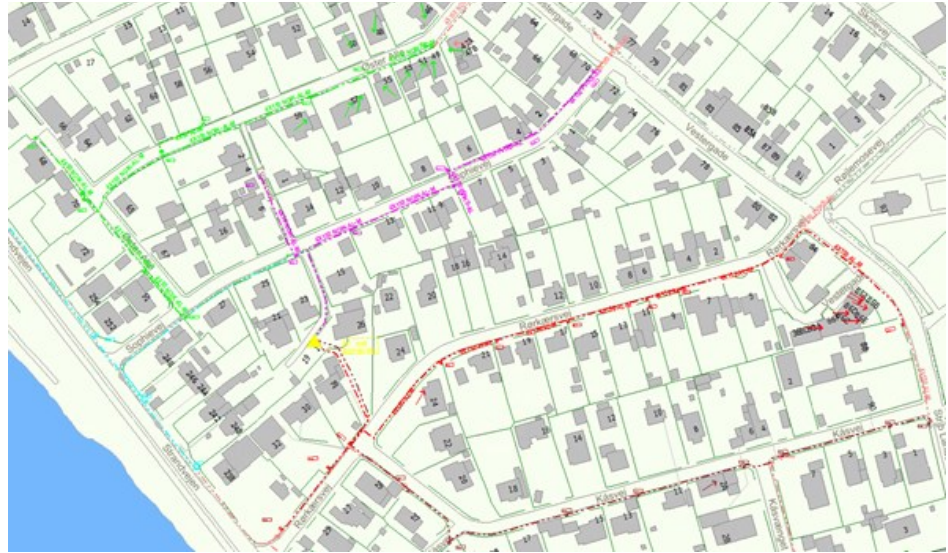

**Figure 7.** The distribution network in Strib, Denmark, used for the case study.

## 4.2. Baseline Consumption Data

An hourly consumption dataset from 2019 was used to represent the baseload, as shown in Figure 8. Although the figure axis is labeled with the simulation year 2025, the underlying data originate from 2019. Because the dataset is limited to the hourly resolution, the simulation assumed that the recorded kilowatt-hour values of each hour equate directly to the load in kilowatts. While short load spikes cannot be fully captured, the comparatively lower magnitude of household baseload relative to EV charging suggests that the overall conclusions remain robust. Photovoltaic systems, heat pumps, and existing EVs were removed from the data to isolate the impact of newly adopted EVs. Future analyses should, however, incorporate these additional DERs for a more comprehensive representation of the evolving ecosystem.

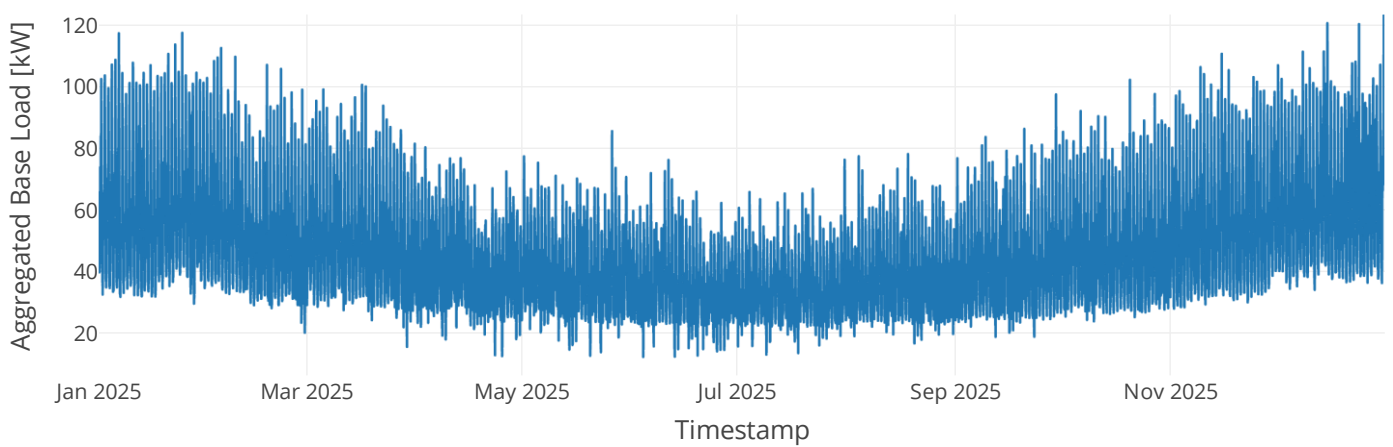

**Figure 8.** The aggregated baseload for the case study.

## 4.3. EV Modeling and Driver Behavior

The EV models and adoption probabilities used here were derived from Danish registration statistics discussed in [65]. The charging power ranged from 7.2 kW to 22 kW, though vehicles capable of 22 kW were restricted by the household limit of about 17.3 kW. Additional simulation inputs, including carbon dioxide emission factors and driving patterns (arrival, departure, and distance traveled), were based on the methods explained in [63]. In the absence of actual arrival and departure data, these behaviors were inferred by identifying significant load increases above the idle load in the morning and afternoon periods, which likely correspond to leaving for and returning from work. When a clear spike was absent, arrival and departure times were randomly assigned within plausible morning and evening intervals. Figure 9 illustrates an example of how arrival and departure times are determined. Driving distances used the probability distribution from [66], derived from large-scale surveys of private vehicle users in Denmark.

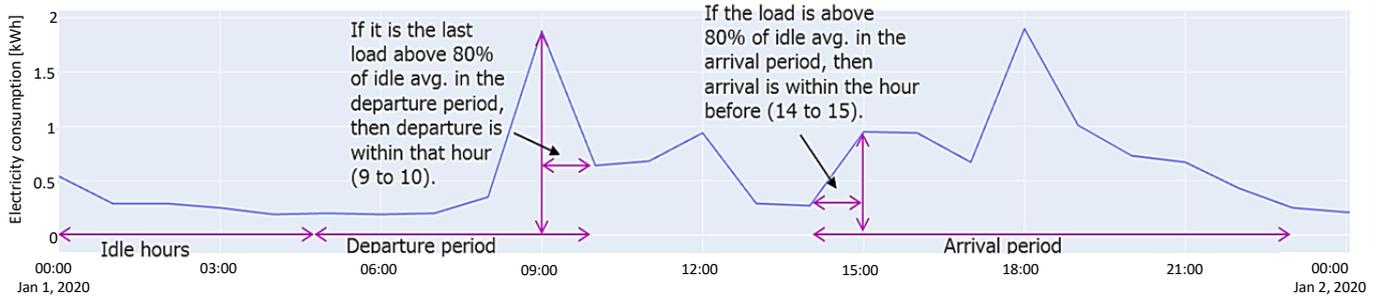

**Figure 9.** Example illustrating how arrival and departure times are inferred [63].

Figure 10 displays the distribution of charging loads for the simulated vehicles, indicating that most users charge at around 11 kW.

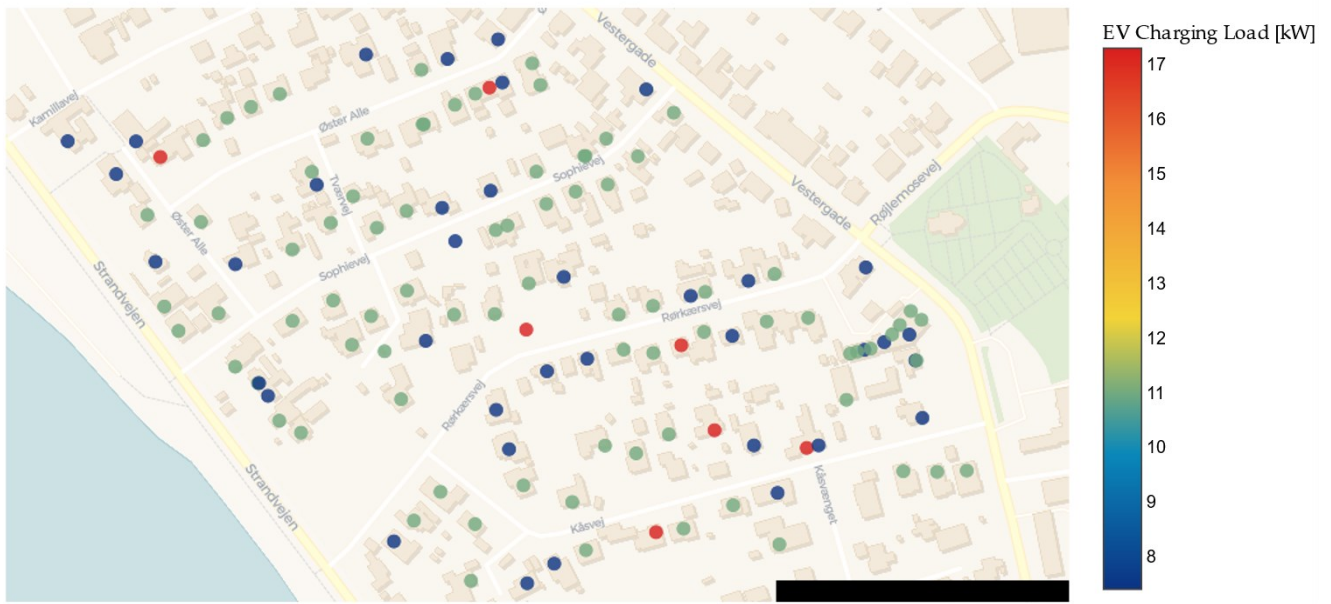

**Figure 10.** The charging loads for the adopted EVs in the grid.

*4.4. Electricity Pricing, Tariffs, and Time Frames*

Day-ahead spot price data from 2024 for DK1 (the price zone containing Strib) were employed in the simulations. The local DSO, Trefor, applies a time-of-use tariff schedule for low-voltage customers, denoted as Tariff Model 3.0, with rates summarized in Table 2. The top portion of Figure 11 shows the annual variability of the spot prices, while the lower portion shows the transition from summer to winter tariffs. Since 2021, Danish electricity consumers can choose hourly billing, prompting many EV charging providers to offer automated "smart charging" services that target the lowest price hours [67–71]. This RTP strategy serves as the default assumption for decentralized charging in the baseline scenario.

**Table 2.** Distribution grid tariffs for C-customers from Trefor, January 2025 [72].

| Hour of Day | Winter [Ore/kWh] | Summer [Ore/kWh] |
|---|---|---|
| 24:00–6:00 | 9.04 | 9.04 |
| 6:00–17:00 | 27.10 | 13.55 |
| 17:00–21:00 | 81.31 | 35.24 |
| 21:00–24:00 | 27.10 | 13.55 |
| Load types | | |
| Low | High | Peak |

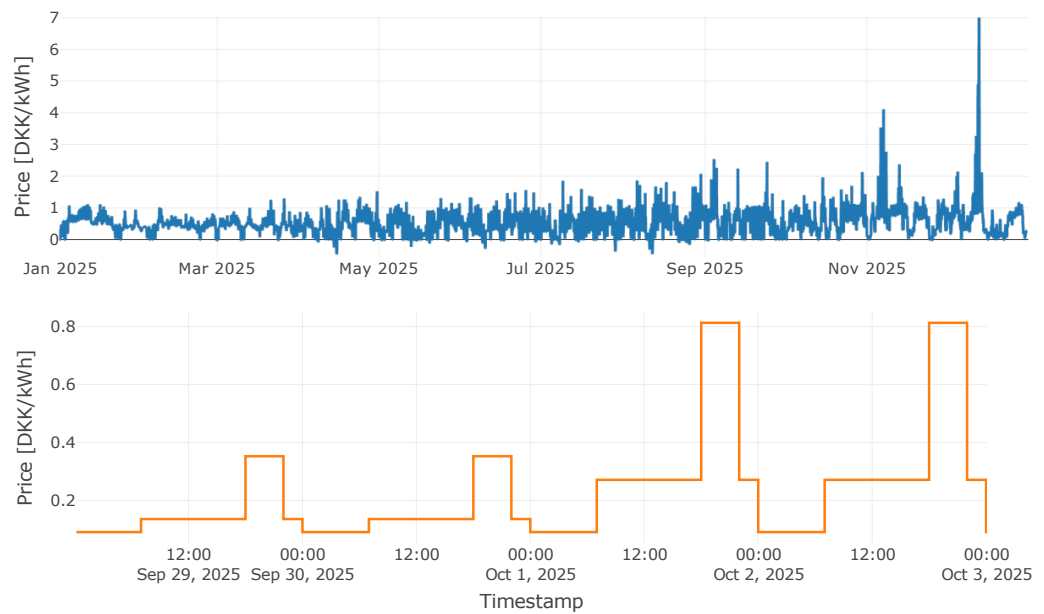

**Figure 11.** (**Top**) Electricity spot prices for 2024 and (**Bottom**) distribution tariffs during summer–winter transition.

### 4.5. Experiment Setup and Scenario Definitions

To evaluate the effectiveness of an aggregator-based charging solution, two primary scenarios were modeled. The baseline scenario represents the current trends of high EV adoption, with all EVs employing decentralized RTP. The second scenario added an aggregator role (as described in the methodology) and assumes full participation in a flexibility service, allowing the aggregator to centrally modify the charging schedules to avoid transformer overload. All inputs, including electricity prices, were otherwise identical.

Table 3 summarizes the experimental parameters. The simulation spans one year (2025) with minute-level data recording. EV adoption was assumed to be 100%, meaning all 126 households own an EV. The only difference between the two scenarios is the presence (or absence) of an aggregator that executes centralized load shifting.

**Table 3.** Experiment setup for baseline and aggregator scenarios.

| Parameter | Value |
|---|---|
| Simulation time | 1 year (2025) |
| Spot price data | 2024 for DK1 |
| Distribution tariff | Tariff Model 3.0 (Trefor 2025) |
| Decentralized charging strategy | Real-time pricing |
| Transformer capacity | 400 kW |
| EV adoption share | 100% = 126 EVs |
| EV charging load | 7.2–17.3 kW |
| Plug-in/out frequency | Once a day |
| Data measurement frequency | 1 min |
| Baseload data | Consumption data for 2019 |
| Driving distance | Data on driving statistics [63] |
| Centralized charging strategy | Variable—based on scenario |

Intermediate cases, such as partial aggregator participation or lower EV adoption, were not explicitly examined in this study but are recognized as important extensions. In real-world settings, compensation incentives could affect user enrollment over time, necessitating iterative modeling of how participation rates change in response to aggregator actions.

*4.6. Key Performance Indicators*

The assessment of each scenario used Key Performance Indicators (KPIs) adapted from [3,63], and the KPIs in the study included the following:

- Total time of overload in hours
- Load factor
- Daily average coincidence factor
- Average charging cost (average of all consumers)
- Average $CO_2$-eq. emissions from charging (average of all consumers)
- DSO Tariff Revenue
- EV users' dissatisfaction

Overloads on the transformer occur whenever the load exceeds the nominal 400 kW capacity (assuming a power factor of 1). While transformers can exceed nameplate ratings briefly without catastrophic damage, the simulation recorded overload any time the power surpasses 400 kW and collected statistics on the overload size, aligning with the recommended guidelines in IEC 60076-7 [73]. Table 4 summarizes the transformer loading limits for small transformers, noting that short-term loads up to 200% may be tolerable in emergencies.

**Table 4.** Recommended current limits for loading beyond small transformer capacities [73].

| Types of Loading | Loading Limits | Definition |
|---|---|---|
| Normal cyclic loading | 150% | "Loading in which a higher ambient temperature or a higher-than-rated load current is applied during part of the cycle, but which, from the point of view of relative thermal ageing rate… is equivalent to the rated load at normal ambient temperature" [73]. |
| Long-time emergency loading | 180% | "Loading resulting from the prolonged outage of some system elements that will not be reconnected before the transformer reaches a new and higher steady-state temperature" [73]. |
| Short-time emergency loading | 200% | "Unusually heavy loading of a transient nature (less than 30 min) due to the occurrence of one or more unlikely events which seriously disturb normal system loading" [73]. |

Additional KPIs evaluate the temporal distribution of load, overall charging costs, carbon emissions, and consumer satisfaction. The load factor (LF) and coincidence factor (CF) follow the International Electrotechnical Commission definitions [74,75], with Equations (1) and (2) computed over each time step *t*:

$$\text{LF}(t) = \frac{\sum_{t=0}^{N_h} C(t)}{\max_t P(t) \cdot N_h} \quad (1)$$

$$\text{CF}(t) = \frac{\max_t P(t)}{\sum_{t=0}^{N_h} \max_t P_i(t)} \quad (2)$$

where $N_h$ is the number of hours in the period, $C(t)$ is the aggregated consumption (kWh), $P(t)$ is the aggregated load (kW), and $P_i(t)$ is the individual load for each consumer. A high LF implies a stable, flattened load profile. A low CF indicates that not all consumers peak at the same time, which is generally beneficial for transformer loading.

Consumer dissatisfaction is measured by Equation (3), which counts the number of instances in which an EV user departs with a state-of-charge below the desired 100% target:

$$\text{Dissatisfactions} = \sum_{T=1}^{N_T} \sum_{EV=1}^{N_{EV}} \begin{cases} 1, & \text{if SoC}_a(EV,T) < \text{SoC}_{EV}(EV,T) \\ 0, & \text{otherwise} \end{cases} \quad (3)$$

where $N_T$ is the total number of daily charging periods, $N_{EV}$ is the total number of EVs, $SoC_a$ is the actual state-of-charge at departure, and $SoC_{EV}$ is the user's desired level.

## 5. Results

The simulation results provide a comparative assessment of the baseline and aggregator scenarios, focusing on transformer loading, user costs, and overall grid stability. In the baseline scenario, high EV penetration leads to frequent transformer overload events, with peak loads significantly exceeding the nominal 400 kW capacity. By contrast, the aggregator-based coordination effectively eliminates overloads by dynamically rescheduling charging sessions, with only a marginal increase in user electricity costs. KPIs such as LF, CF, and DSO tariff revenues are analyzed to quantify the benefits of centralized charging coordination. Additionally, a cost–benefit analysis highlights the economic advantages of aggregator-based load shifting over costly infrastructure reinforcements.

### 5.1. Baseline Scenario: Full EV Adoption and Real-Time Pricing

The first scenario assumes that all 126 households in the network own an EV and employ a decentralized RTP charging strategy. Figure 12 illustrates the aggregated load on the transformer over one simulated year, where the red dashed line marks the 400 kW capacity. The simulation reveals 587 h of overload throughout the year. Notably, the summer months show intermittent "gaps" in simultaneous charging peaks due to lower electricity prices mid-day from solar power.

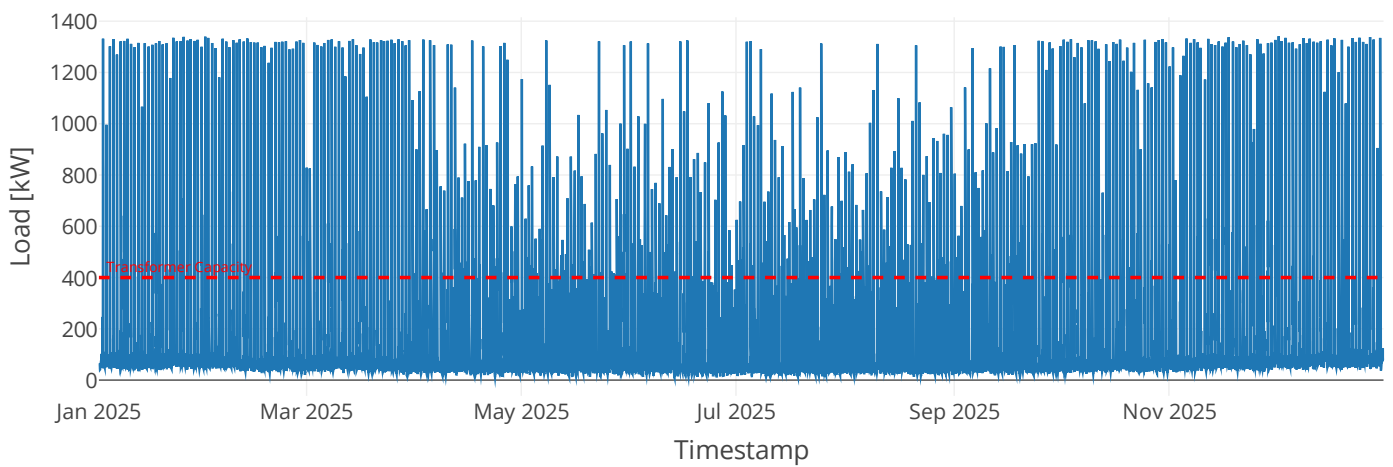

**Figure 12.** The aggregated transformer load for the baseline scenario, with the red line indicating the 400 kW capacity.

Further insights into the interplay between electricity prices, charging behavior, and overloads are provided in Figure 13, which depicts the total electricity price (comprising spot price plus distribution tariff) alongside the number of actively charging EVs. During summer, mid-day prices are often the lowest, causing many EVs that arrive before the 17:00 peak tariff to begin charging immediately and thus avoid nighttime charging congestion.

This behavior explains the intermittent "gaps" in simultaneous charging peaks during the summer months in Figure 12. Although this price-driven behavior can reduce the nighttime load, the daily load profile can still exceed the transformer's nominal capacity.

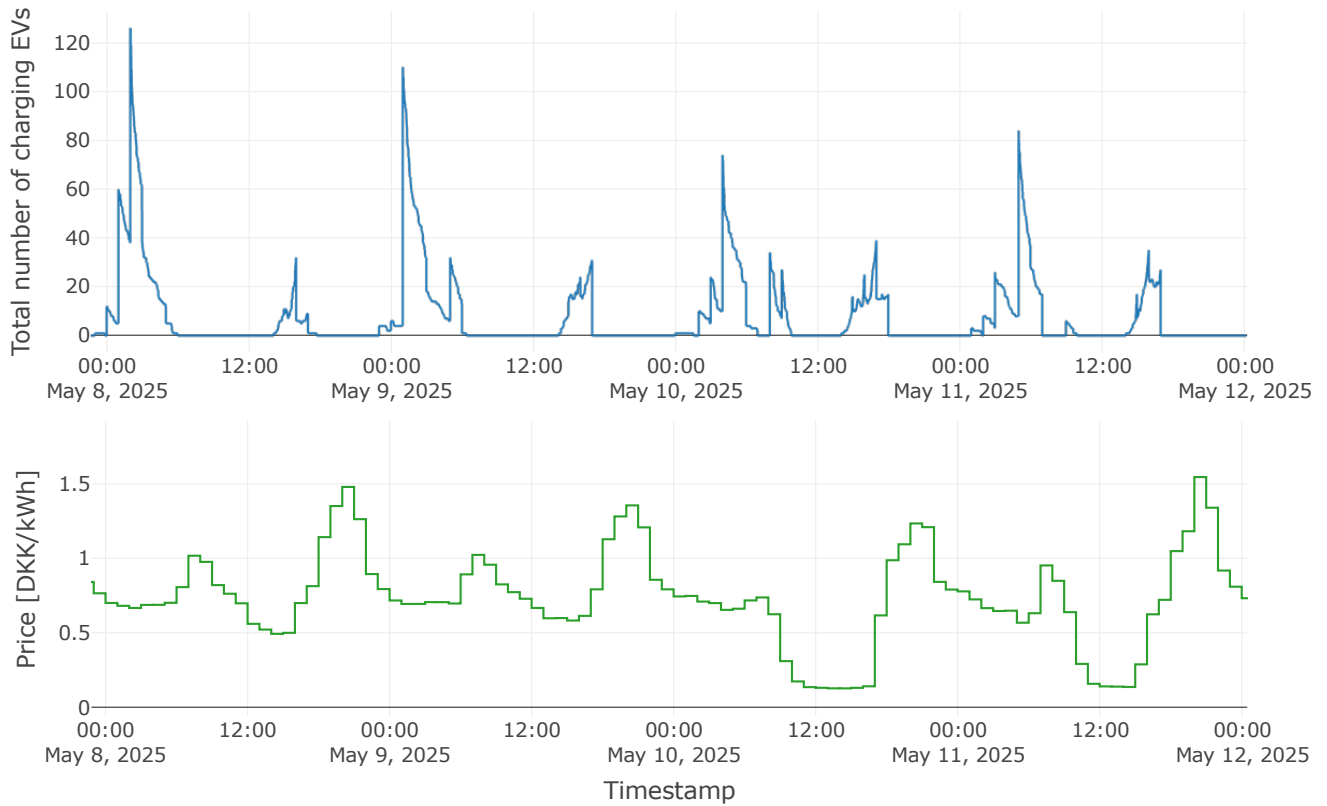

**Figure 13**. (**Top**) Number of actively charging EVs; (**Bottom**) total electricity price (spot price + tariff).

The severity of overload is characterized in Figure 14, which compares the observed load to the recommended loading limits in Table 4. Over 95 h surpass 200% of the nominal capacity (800 kW), categorized as critical load and with a maximum peak of 1342 kW. Although short-time emergency overloading is theoretically permitted [73], these extreme peaks suggest substantial stress on the transformer.

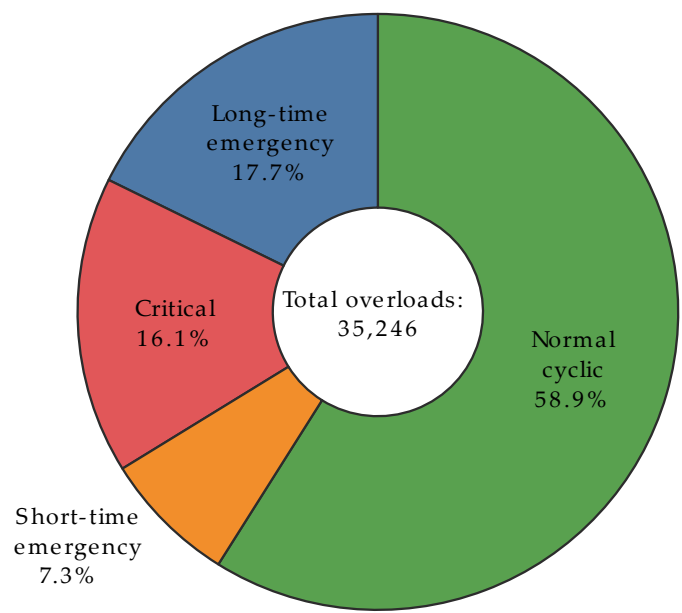

**Figure 14.** Frequency of overload events categorized by overload level, summed in minutes per year.

Figure 15 zooms in on a single 24 h charging period to illustrate how EVs cluster in the cheapest hours. The top sub-figure displays the aggregated transformer loading along with the number of charging EVs and the total electricity price. The bottom sub-figure offers a closer view of part of that period. Figure 16 shows a heat map, illustrating the distribution of the charging EVs for the same period. Most vehicles begin charging in the lowest-priced hour, and some finish charging before peak tariffs. A single EV is observed charging during an expensive hour (23:00–23:21), triggered by its constrained departure time (05:08) and low initial state of charge.

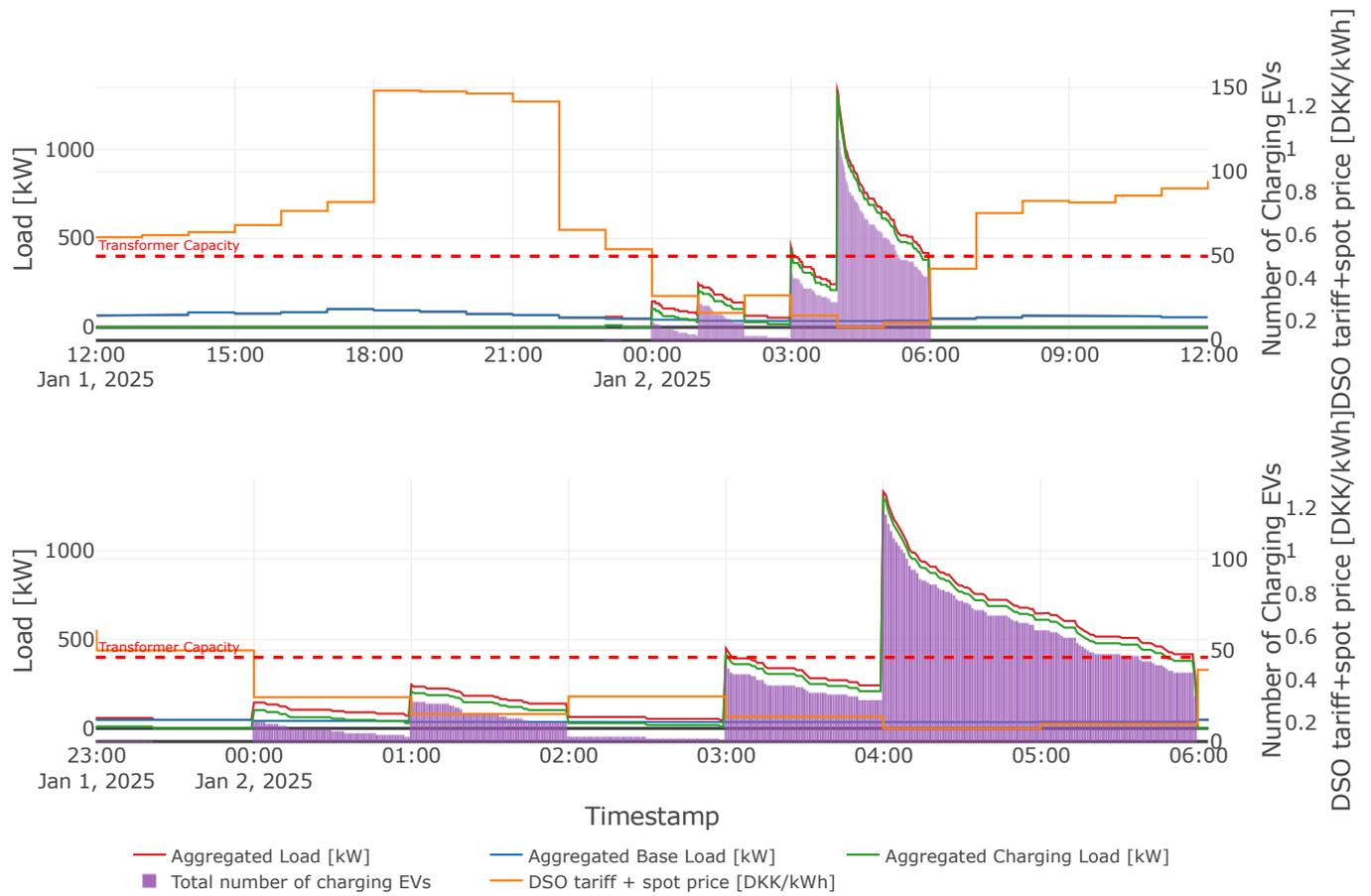

**Figure 15.** The baseline scenario. (**Top**) One day's loading profile with the charging EV count and electricity price. (**Bottom**) A zoomed portion of the same day.

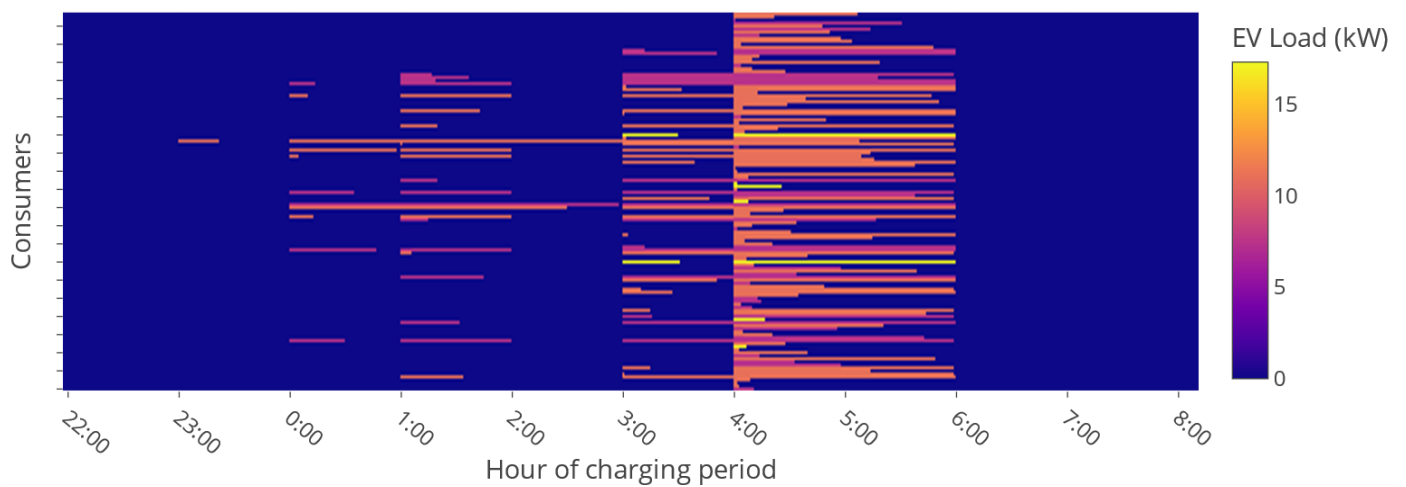

**Figure 16.** Baseline EV charging heatmap for January 1st to 2nd, 2025.

Table 5 presents the KPIs for the baseline scenario. A total of 587 overload hours occurs, highlighting the scale of potential transformer stress. The LF is low (0.089), indicating a highly fluctuating load curve, while the CF is close to 1 (0.826), reflecting that most users peak simultaneously. The average charging cost remains relatively low (0.48 DKK/kWh) due to strategic charging at cheaper hours. The $CO_2$-equivalent emissions factor is 0.2175 kg/kWh, and the DSO tariff revenue sums to approximately 179,500 DKK.

**Table 5.** KPIs for the baseline scenario.

| Total Time of Overload [Hours] | Load Factor | Daily Average Coincidence Factor | Average Charging Cost [DKK/kWh] | Average $CO_2$-eq. Emissions from Charging [kg/kWh] | DSO Tariff Revenue [DKK] | EV Users' Dissatisfaction |
|---|---|---|---|---|---|---|
| 587 | 0.089 | 0.826 | 0.48 | 0.2175 | 179.5k | 0 |

In practice, harmonizing aggregator-based EV management with recognized standards can simplify implementation and interoperability. Relevant protocols include ISO 15118 for vehicle–grid communication [76], IEC 61850 for utility automation [77], OCPP (Open Charge Point Protocol) [78], IEEE 2030.5 [79], and ENTSO-E Network Codes [80]. Adherence to such standards can help ensure compatibility across different platforms and devices.

*5.2. Aggregation Scenario: Centralized Coordination to Avoid Overloads*

The second scenario introduces an aggregator role, managed by the DSO that is responsible for shifting the EV load away from overloaded hours. Figure 17 shows that overload events are effectively eliminated under this centralized strategy.

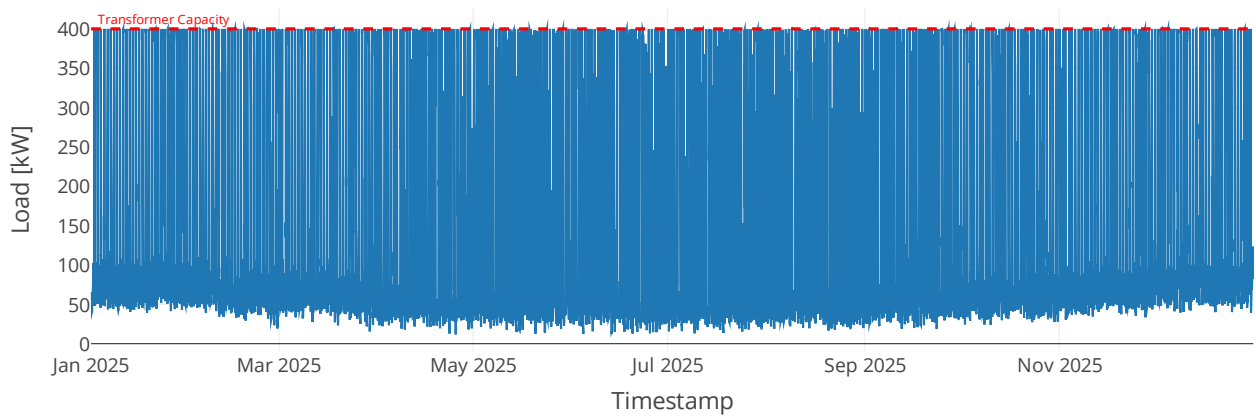

**Figure 17.** The aggregated transformer load in the aggregation scenario.

Figures 18 and 19 focus on the first day's loading pattern (from noon to noon) to illustrate how the aggregator reallocates charging. Compared to the baseline, charging is distributed across more time slots to keep the load under the 400 kW threshold. Some shifts move EV charging into hour 6, which has a higher price than hour 2 but remains within capacity limits. This cost trade-off underscores the aggregator's priority to alleviate overload first, followed by finding the cheapest feasible hour.

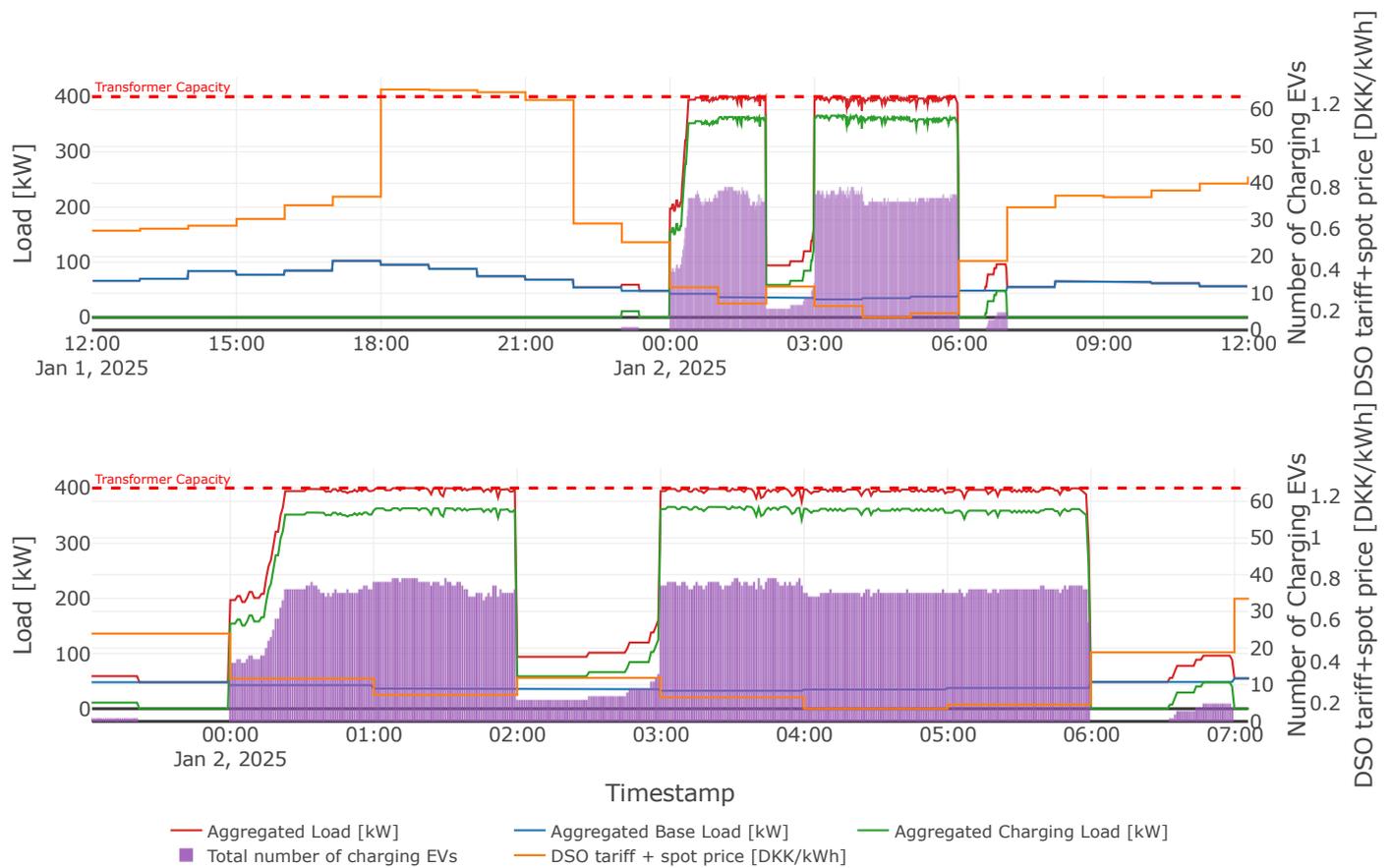

**Figure 18.** The aggregation scenario. (**Top**) A single day's load profile with the charging EV count and electricity prices. (**Bottom**) A zoomed portion of the same day.

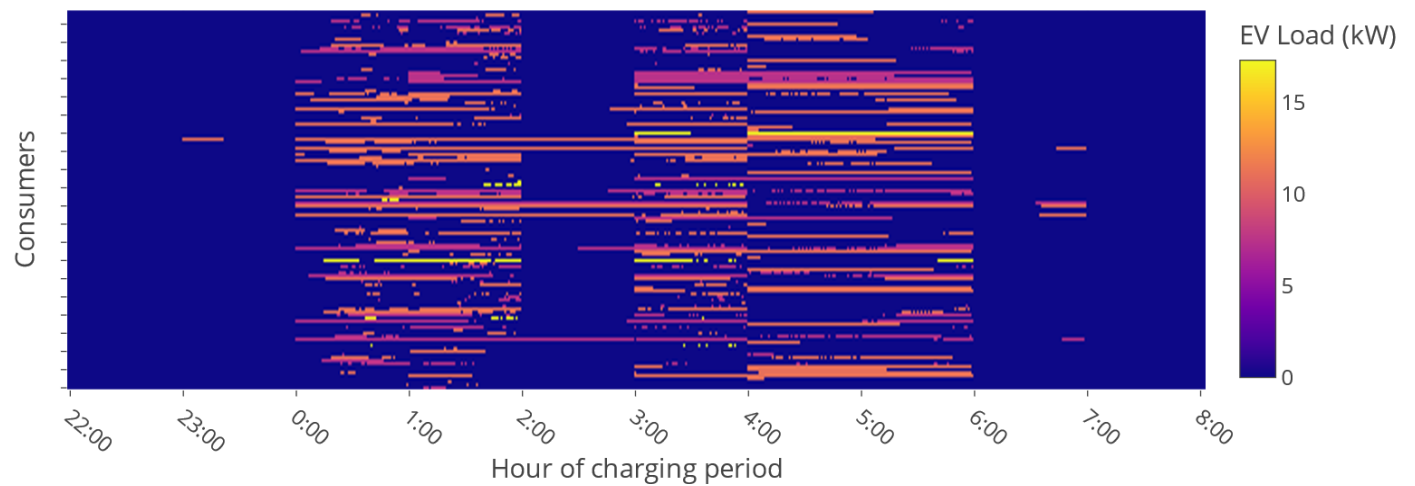

**Figure 19.** Aggregation scenario EV charging heatmap for January 1st to 2nd, 2025.

Figure 20 compares the number of charging EVs for both scenarios over the entire year (top) and on a single day (bottom). Under the aggregator strategy, simultaneous charging seldom exceeds 40 EVs, whereas the baseline allows up to 126 EVs to initiate charging in the same period. This difference demonstrates the aggregator's capability to limit concurrency and reduce peak load. However, it also implies that if more than approximately 30% of vehicles lack scheduling flexibility, overloads might reoccur under high adoption levels.

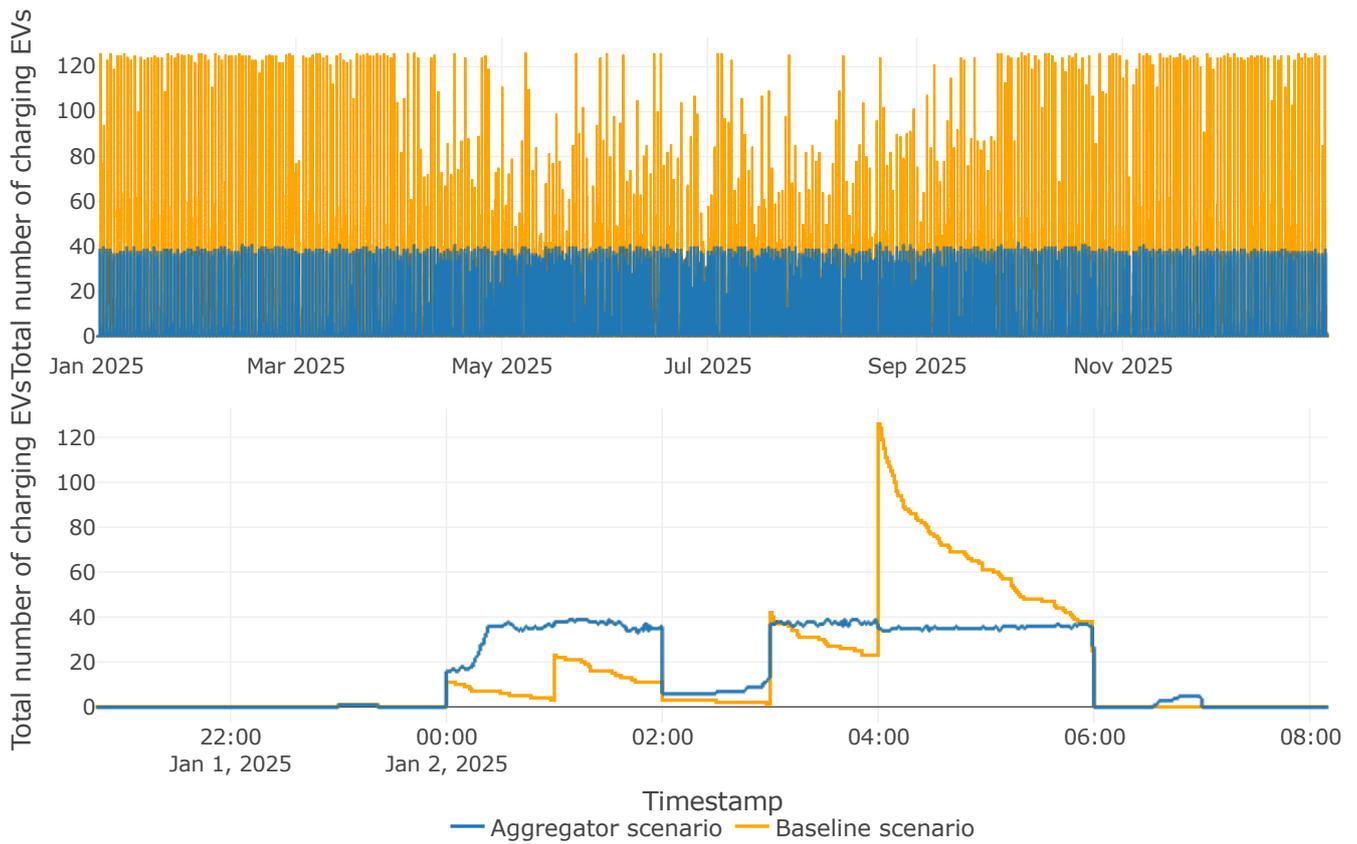

**Figure 20.** (**Top**) The yearly number of actively charging EVs for both scenarios. (**Bottom**) A comparison for a single day.

Table 6 summarizes the KPIs for the aggregation scenario. The transformer overload drops to zero, the load factor increases to 0.299, and the coincidence factor declines to 0.301. The average charging cost rises slightly to 0.489 DKK/kWh (a 1.8% increase from the baseline), and emissions rise by 0.2%, reflecting some shifts into hours with marginally higher carbon intensity. The revenue for the DSO remains effectively unchanged, indicating that although individual hours might change, the overall distribution tariff structure is preserved.

**Table 6.** KPIs for the aggregation scenario.

| Total Time of Overload [Hours] | Load Factor | Daily Average Coincidence Factor | Average Charging Cost [DKK/kWh] | Average $CO_2$-eq. Emissions from Charging [kg/kWh] | DSO Tariff Revenue [DKK] | EV Users' Dissatisfaction |
|---|---|---|---|---|---|---|
| 0 | 0.299 | 0.301 | 0.489 | 0.218 | 179.5k | 0 |
| Percentage difference to baseline | | | | | | |
| −100% | 235% | −63.6% | 1.8% | 0.2% | - | - |

Relative changes from the baseline are shown in the lower part of Table 6: –100% overload, a 235% improvement in the LF, and a 63.6% decrease in the CF. Although charging costs and emissions increase slightly, these effects remain modest.

The observed improvements in the LF and CF reflect a more efficient and distributed use of grid capacity over time. A higher load factor indicates that the transformer is utilized more consistently, reducing idle periods and spreading demand more evenly. The sharp decline in the coincidence factor suggests that EVs are charging at fewer overlapping times, alleviating peak-related stress. From a system operation perspective, these effects reduce the likelihood of thermal overloads, extend asset lifetimes, and defer the need for capital-intensive upgrades. Furthermore, the preservation of DSO revenue despite these changes implies that load shifting can be effectively implemented within the existing pricing structure of Tariff Model 3.0, particularly by utilizing the lowest-cost period between 00:00 and 06:00.

The aggregator also pays a total compensation of about DKK 6020 to users, primarily directed toward high-capacity EVs, as visualized in Figure 21. Vehicles with larger charging powers tend to complete charging faster, thus exhibiting higher "laxity" and being more frequently selected for load shifting. This total compensation is small compared to infrastructure upgrade costs.

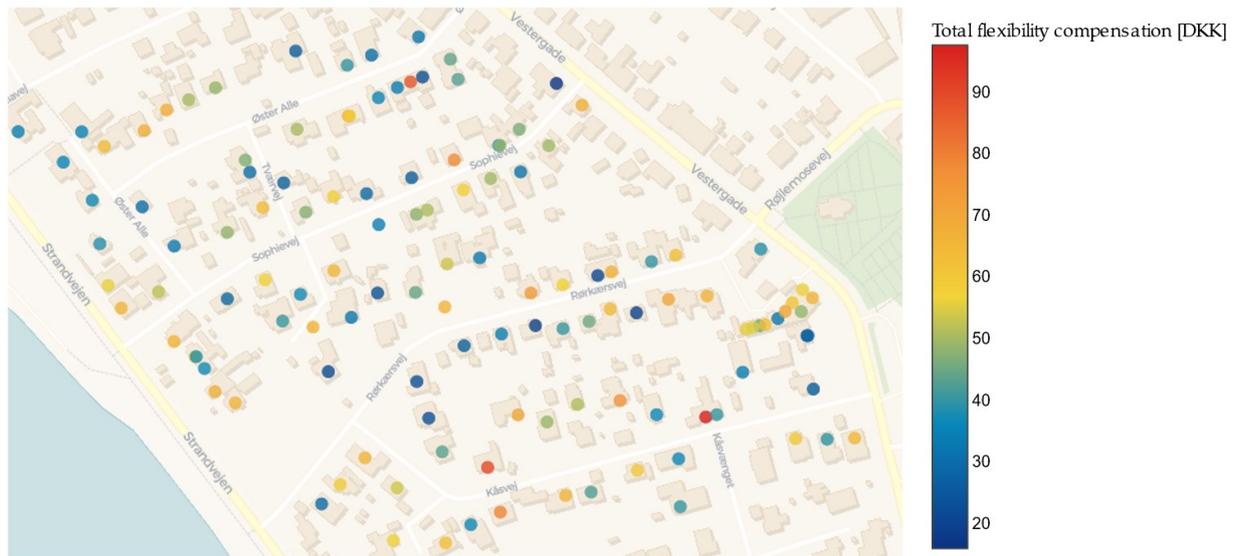

**Figure 21.** The compensation paid by the aggregator to each EV user.

*5.3. Cost–Benefit Analysis of Aggregation vs. Transformer Upgrades*

Comparisons of transformer reinforcement costs help contextualize the aggregator's financial impact. According to estimates from the Danish Energy Agency, upgrading transformer capacity can range from about DKK 115.8k (reinforcement) to DKK 679.8k (new station) when the network load reaches 1350 kW [81]. Vendor prices for a new 400 kVA transformer are around DKK 240k [82], while a 1600 kVA transformer costs about DKK 589k [83]. These figures exclude installation expenses. Given an annual aggregator compensation of only DKK 6000, the payback for a DKK 115.8k upgrade would stretch to nearly two decades, and a more expensive new station could take over a century. Hence, aggregator-based coordination presents a cost-effective alternative that can defer upgrades while adapting to evolving load patterns from future DERs.

## 6. Discussion

The simulation results presented in this paper confirm that a simple, rule-based aggregator design can virtually eliminate transformer overload in a highly penetrated EV environment. By assigning each EV a "laxity" measure—an indicator of how freely its charging window can be shifted—the aggregator relieves the local 400 kVA transformer from frequent overload events. In the baseline scenario, with all EVs following decentralized RTP, the transformer was overloaded for hundreds of hours each year. Once the aggregator was introduced, the overload hours fell to zero, demonstrating how even modest load-shifting actions, distributed over many participants, can significantly reduce peak demand. Compared to more complex scheduling algorithms, this rule-based approach combines ease of implementation with strong performance.

One distinctive outcome of the study is the demonstration that compensating EV users for any additional costs incurred by rescheduling their charging remains considerably less expensive than upgrading or reinforcing the transformer. While some EVs experience marginally higher electricity prices due to load shifts, the total annual compensation payment is on the order of a few thousand Danish kroner—an amount that is minimal compared to the tens or hundreds of thousands often required for expanding physical infrastructure. This finding positions aggregator-based solutions as credible alternatives to capacity upgrades, especially in markets and regulatory settings that aim to minimize infrastructure expenditure.

These observations align with past research, which shows that the centralized coordination of EV charging can mitigate grid congestion and potentially offer ancillary services [9–11,25–29]. However, unlike optimization-based or machine-learning solutions that often require extensive real-time data processing [12,13,37], the present work highlights a computationally light technique. The aggregator leverages locally available information on time windows and price signals, supplemented by a straightforward laxity criterion. Such simplicity increases the feasibility of real-world deployment by DSOs that typically lack robust forecasting or optimization capacities [34,35].

Cost analyses in prior studies typically emphasize the economic efficiency of dynamic tariffs or advanced control strategies in lowering consumer bills [14,15]. In contrast, the present paper treats aggregator-driven load shifting primarily as an alternative to capital-intensive grid reinforcements. This perspective resonates with [36,45], who underscore the potential of localized flexibility markets to defer or obviate infrastructure upgrades. Moreover, while earlier works have recognized the conceptual promise of aggregator services, most have not systematically compared aggregator compensation outlays with transformer replacement or reinforcement costs. The simulations reported here fill this gap, revealing that aggregator payments are a fraction of typical upgrade budgets.

The multi-agent simulation framework, augmented by a business ecosystem modeling approach [20,21], is another distinguishing factor. Previous aggregator studies often rely on either theoretical models with limited real-world detail or large-scale market simulations that overlook individual decision-making processes. By embedding each EV user, the aggregator, and the DSO as separate agents, the model captures micro-level charging decisions and macro-level capacity constraints. This hybridization offers a granular understanding of how minor scheduling changes accumulate into significant peak load reductions, bridging a methodological gap noted in earlier reviews of EV aggregator research [9,10].

By applying laxity as a practical scheduling parameter, the study also builds upon earlier scheduling algorithms that prioritized tasks based on flexibility [16]. Yet, most prior work did not focus on large-scale, real-time aggregator solutions explicitly incorporating capacity constraints at the distribution transformer. By demonstrating near-complete overload mitigation through laxity-based sorting, this paper provides a straightforward decision rule that DSOs could adopt to identify which EVs to shift during congested periods without requiring advanced optimization engines.

The aggregator concept can be extended to manage other DERs, including residential photovoltaic systems, home batteries, and heat pumps. As [25,29] indicate, harnessing multiple types of flexibility can further flatten demand peaks and help align local consumption with renewable generation. However, each additional DER type introduces more complex interactions, real-time forecasting needs, and communication overheads. The present research establishes a foundation for such integration by underscoring how agent-based modeling can incorporate diverse household technologies. Future studies can explore how rule-based aggregator models might scale to control not just EV loads but also controllable DER devices under the same framework, potentially improving the network's load factor and stability even further.

In focusing on the transformer-level domain, this paper offers a new angle on aggregator research. Previous work has often centered on wholesale market participation or broad distribution planning without delving deeply into local overload dynamics. By merging a business ecosystem modeling approach [20,21] with an agent-based simulation, this study unifies the microeconomics of user decisions and the macroeconomics of grid constraints. The laxity metric, while not novel in its scheduling theory [16], has not been widely explored as a straightforward aggregator tool for distribution system operators. Its demonstrated efficacy here underscores its potential as a low-complexity alternative to more advanced optimization or machine learning methods [12,13,37].

In addition, the explicit comparison between aggregator compensation costs and hardware reinforcement costs provides DSOs with actionable insights. Rather than merely lowering consumer bills, this aggregator mechanism focuses on deferring or avoiding infrastructure upgrades. Thus, the results respond directly to a key policy and operational question: how to reconcile large-scale EV adoption with minimal strain on distribution assets.

A key strength underpinning the relevance of this study lies in the adaptability and scalability inherent in the simulation framework design. The adoption of an agent-based modeling platform facilitates a modular architecture where entities like EVs, households, and the aggregator function as independent agents. This modularity inherently supports scalability, allowing the model to accommodate larger systems or varying network structures with relative ease. More critically for broader applicability, the framework is highly adaptable through its parameterization to other regions that have electricity markets, an unbundled electricity system, and a regulatory framework similar to Denmark. Key inputs that define the operational environment—such as regional baseload consumption patterns, specific DSO tariff structures, and statistical distributions governing EV availability and charging needs —can be readily modified without altering the core simulation logic. This design choice makes the framework a versatile tool that is capable of being repurposed to analyze the effectiveness of the proposed laxity-based aggregation strategy under different socio-technical conditions beyond the Danish suburban context examined here.

One important limitation is the assumption of full and passive user participation: once enrolled, all EV owners accept aggregator decisions and maintain consistent charging patterns. Realistically, users may opt in or out based on evolving costs, incentives, or preferences, necessitating partial adoption models that track the aggregator's dynamic enrollment [45,47]. There is also scope to refine compensation mechanisms, shifting from a flat cost offset to more nuanced, market-based valuations of flexibility, as suggested by [46,49]. Extending the aggregator's scope to multiple low-voltage feeders or coordinating several aggregators in the same region also merits investigation, given that different aggregator strategies might compete or conflict, potentially affecting overall grid performance.

Moreover, this study does not explicitly account for uncertainties such as random events, data latency, grid voltage, and EV battery health concerns. Addressing these aspects may require scenario-based analyses that combine rule-based aggregator models with real-time sensing and forecasting, bridging the gap between high-level theoretical frameworks and day-to-day grid operations. Lastly, evaluating how aggregator solutions interface with standard protocols like ISO 15118, IEC 61850, and OCPP remains important for commercial scalability [76–80].

In summary, the key achievement of this work lies in operationalizing a rule-based aggregator strategy that balances effectiveness, simplicity, and cost-efficiency. The use of a laxity measure enables prioritization without advanced optimization, while the agent-based framework captures both user behavior and grid constraints with high fidelity. The economic analysis reinforces the practical value of this approach, showing that compensation costs are negligible compared to hardware upgrades. Together, these results support a shift in focus from consumer cost savings toward infrastructure deferral—highlighting a DSO-oriented use case for aggregator-based flexibility that is both technically feasible and economically rational.

## 7. Conclusions

This study has demonstrated that a rule-based aggregator, guided by a laxity-based scheduling mechanism, effectively eliminates transformer overload events in a low-voltage distribution grid hosting 126 electric vehicles. In contrast to the baseline scenario, where more than 500 overload hours occur each year, the aggregator scenario removes all overload episodes. This outcome is achieved at an aggregate compensation cost that remains far below the capital expenses of upgrading or replacing the existing 400 kVA transformer. The agent-based simulation design highlights how simple, real-time interventions, aggregated over many end-users, can result in major peak reductions.

From a regulatory and operational standpoint, the aggregator approach suits DSOs operating under constraints that limit the adoption of fully dynamic tariffs [4,5]. By handling congestion via a targeted, rule-based method, DSOs can reduce the complexity of user engagement. Yet, real-world deployments will require careful management of partial user participation, communication protocols, and the interplay with other DERs. Exploring these contingencies, including how aggregator services might interface with local flexibility market mechanisms [45–47], forms a critical step toward commercial viability.

Future research can extend the aggregator framework to account for partial or intermittent participation, incorporate advanced demand forecasting, and coordinate with multiple aggregators across adjoining low-voltage networks. Such investigations will shed light on competition or synergy among aggregator strategies, especially when diverging objectives coexist. Quantifying the impact of market-based flexibility pricing on user enrollment and aggregator profit could further illuminate the economic sustainability of these solutions. Finally, investigating multi-modal aggregator control—wherein heat pumps, battery storage, and EVs are all actively managed—could reveal additional gains in grid stability and peak reduction.

In conclusion, this study's agent-based evaluation of a simple, laxity-based aggregator demonstrates that local transformer overload can be eliminated at a fraction of the cost of capacity upgrades. The findings confirm that, given appropriate regulatory and market frameworks, DSOs can employ aggregators as an indispensable tool to meet the rising EV adoption demands, sustaining grid reliability and cost-effectiveness in the face of rapidly evolving energy ecosystems.

**Author Contributions:** Conceptualization, K.C., Z.G.M., and B.N.J.; Data curation, K.C.; Formal analysis, K.C.; Funding acquisition, Z.G.M. and B.N.J.; Investigation, K.C.; Methodology, K.C., Z.G.M., and B.N.J.; Project administration, Z.G.M. and B.N.J.; Resources, Z.G.M. and B.N.J.; Software, K.C.; Supervision, Z.G.M. and B.N.J.; Validation, K.C.; Visualization, K.C.; Writing—original draft, K.C.; Writing—review and editing, Z.G.M. and B.N.J. All authors have read and agreed to the published version of the manuscript.

**Funding:** This research is part of the project "Digital Energy Hub", funded by the Danish Industry Foundation, and part of the project "The Danish Participation in IEA ES Task 43—Storage for renewables and flexibility through standardized use of building mass", funded by EUDP (case number: 134232-510227).

**Institutional Review Board Statement:** Not applicable

**Informed Consent Statement:** Not applicable

**Data Availability Statement:** This study utilized publicly available data, including historical baseload consumption records, EV registration statistics, driving patterns, electricity market prices, and transformer cost estimates from official reports and vendor quotations. Simulation-generated data supporting the findings of this research may be available from the corresponding author upon reasonable request. Data provided by distribution system operators are confidential and not publicly shareable.

**Conflicts of Interest:** The authors declare no conflicts of interest.

## Abbreviations

The following abbreviations are used in this manuscript:

| | |
|---|---|
| EV | Electric Vehicle(s) |
| DSO | Distribution System Operator(s) |
| kVA | Kilovolt-Ampere |
| DKK | Danish Kroner |
| DER | Distributed Energy Resource(s) |
| VPP | Virtual Power Plant(s) |
| V2G | Vehicle-to-Grid |
| AI | Artificial Intelligence |
| LF | Load Factor |
| CF | Coincidence Factor |
| OCPP | Open Charge Point Protocol |
| IEC | International Electrotechnical Commission |
| ISO | International Organization for Standardization |
| IEEE | Institute of Electrical and Electronics Engineers |
| ENTSO-E | European Network of Transmission System Operators for Electricity |
| RTP | Real-Time Pricing |
| SoC | State-of-Charge |
| LFMs | Local Flexibility Markets |
| DK1 | Danish Electricity Price Zone 1 |